\documentclass[12pt]{article}

\usepackage{graphicx}
\usepackage{fancybox}
\usepackage{amsmath}
\usepackage{amssymb}
\usepackage{latexsym}
\usepackage{epsfig}

\newcommand{\Om}{\Omega}
\newcommand{\om}{\omega}
\newcommand{\al}{\alpha}
\newcommand{\ep}{\epsilon}

\newcommand{\la}{\lambda}
\newcommand{\La}{\Lambda}

\newcommand{\tR}{\widetilde{\mbox{R}}}
\newcommand{\sNS}{\msc{NS}}
\newcommand{\stNS}{\widetilde{\msc{NS}}}
\newcommand{\sR}{\msc{R}}
\newcommand{\stR}{\widetilde{\msc{R}}}

\newcommand{\del}{\partial}

\newcommand{\lb}{\lbrack}
\newcommand{\rb}{\rbrack}

\newcommand{\msc}[1]{\mbox{\scriptsize #1}}
\newcommand{\dsp}{\displaystyle}

\newcommand{\bc}{\Bbb C}
\newcommand{\br}{\Bbb R}
\newcommand{\bz}{\Bbb Z}

\newcommand{\bsz}{\Bbb Z}
\newcommand{\bsr}{\Bbb R}

\newcommand{\cN}{{\cal N}}

\newcommand{\cF}{{\cal F}}

\newcommand{\cC}{{\cal C}}
\newcommand{\cQ}{{\cal Q}}
\newcommand{\cD}{{\cal D}}

\newcommand{\cZ}{{\cal Z}}

\newcommand{\tJ}{\tilde{J}}

\newcommand{\tj}{\tilde{j}}

\newcommand{\hchi}{\widehat{\chi}}

\newcommand{\hc}{\hat{c}}
\newcommand{\hF}{\widehat{F}}
\newcommand{\hf}{\widehat{f}}

\newcommand{\hH}{\widehat{H}}
\newcommand{\hh}{\widehat{h}}
\newcommand{\hG}{\widehat{G}}

\newcommand{\Th}[2]{\Theta_{#1,#2}}
\renewcommand{\th}{{\theta}}

\newcommand{\ch}[2]{\mbox{ch}^{#1}_{#2}}

\newcommand{\chd}{\mbox{ch}_{\msc{dis}}}

\newcommand{\hchd}{\widehat{\mbox{ch}}_{\msc{dis}}}

\newcommand{\chid}{\chi_{\msc{dis}}}

\newcommand{\hchid}{\widehat{\chi}_{\msc{dis}}}

\newcommand{\tr}{\mbox{Tr}}

\newcommand{\tpsi}{\tilde{\psi}}

\newcommand{\erf}{\mbox{Erf}}

\newcommand{\sgn}{\mbox{sgn}}

\newcommand{\nn}{\nonumber\\}

\renewcommand{\Im}{{\rm Im}}

\def\boxit#1{\vbox{\hrule\hbox{\vrule\kern8pt
\vbox{\hbox{\kern8pt}\hbox{\vbox{#1}}\hbox{\kern8pt}}
\kern8pt\vrule}\hrule}}
\def\mathboxit#1{\vbox{\hrule\hbox{\vrule\kern8pt\vbox{\kern8pt
\hbox{$\displaystyle #1$}\kern8pt}\kern8pt\vrule}\hrule}}

\newcommand{\any}{{}^{\forall}}

\renewcommand{\mod}{\, \mbox{mod} ~ }

\newcommand {\eqn}[1]{(\ref{#1})}

\makeatletter
\@addtoreset{equation}{section}
\def\theequation{\thesection.\arabic{equation}}
\makeatother

\begin{document}

\begin{titlepage}
 \
 \renewcommand{\thefootnote}{\fnsymbol{footnote}}
 \font\csc=cmcsc10 scaled\magstep1

 \baselineskip=20pt
 
\begin{center}

{\bf \Large  
Non-Compact SCFT and  \\
Mock Modular Forms

}

 
\vskip 2cm

\vskip 8mm
\noindent{ \large Yuji Sugawara}\footnote{\sf ysugawa@se.ritsumei.ac.jp}
\\

\medskip

 {\it Department of Physical Sciences, 
 College of Science and Engineering, \\ 
Ritsumeikan University,  
Shiga 525-8577, Japan}
 

\end{center}

\bigskip


\begin{center}

{\it Dedicated to the memories of Prof. Tohru Eguchi}

\end{center}

\bigskip

\bigskip

\begin{abstract}

One of interesting issues in two-dimensional superconformal field theories 
is the existence of anomalous modular transformation properties 
appearing in some non-compact superconformal models, corresponding to the `mock modularity' in mathematical literature. 
I review a series of my studies on this issue in collaboration with T. Eguchi, 
mainly focusing on the papers \cite{ES-NH,ES-nhP,ES-N=4L}.

\end{abstract}



\setcounter{footnote}{0}
\renewcommand{\thefootnote}{\arabic{footnote}}

\end{titlepage}

\baselineskip 18pt

\vskip2cm 


\section{Introduction and Summary}

~

Two-dimensional superconformal field theories have been a central subject in the study of string theory for a long time. 
One of intriguing issues is the existence of anomalous modular transformation properties 
in some non-compact or non-rational superconformal models. 
This is called the `mock modularity' in mathematical literature. 
Namely, non-trivial mixtures of discrete and continuous spectra often emerge  
under the modular S-transformation, 
which make it difficult to assure modular invariance in a simple manner. 
Such an anomalous modular behavior has been first observed for
the massless (BPS) characters of $\cN=4$ superconformal algebra (SCA) \cite{ET}.  
Similar behavior appears in the discrete (BPS) characters 
in the $\cN=2$ Liouville theory or the $SL(2,\br)/U(1)$-supercoset model  
\cite{ES-L,ES-BH}
(see also \cite{Odake,Miki}).




The mock modularity is well expressed in terms of the following meromorphic function
often called the `Appell-Lerch sum' \cite{Zwegers,STT}; 
\begin{equation}
f^{(k)}(\tau,z) 
:= \sum_{n\in \bsz} \frac{q^{kn^2} y^{2kn}}
{1-yq^n},
\hspace{1cm} 
\left(k \in \bz_{>0}\right)
\label{Appell}
\end{equation}
which roughly corresponds to the discrete representations of ($\cN \geq 2$) SCA. 
Its modular property is summarized as follows;
\begin{align}
& f^{(k)}\left(-\frac{1}{\tau}, \frac{z}{\tau}\right)
= \tau e^{2 \pi i k \frac{  z^2}{\tau}}\,
\left[ f^{(k)}(\tau,z) - \frac{i}{\sqrt{2 k}}\, \sum_{m\in \bz_{2k}}\,
\int_{\br+i0} dp' \, \frac{q^{\frac{1}{2}p^{'2}}}
{1-e^{-2\pi \left(\frac{p'}{\sqrt{2k}}+i\frac{m}{2k}\right)}}\,
\Th{m}{k}(\tau,2z)
\right],
\label{S Appell}
\\
& f^{(k)}\left(\tau+1, z\right) = f^{(k)}\left(\tau, z \right).
\label{T Appell}
\end{align}
The inhomogeneous term in the R.H.S of \eqn{S Appell} exhibits the mixing of the discrete and continuous spectra 
mentioned above. 
If it would be absent, the function $f^{(k)}(\tau,z)$ \eqn{Appell} would get a (weak) Jacobi form\footnote
   {A basic summary of the (weak) Jacobi form is given in Appdendix A. See \cite{Eich-Zagier} for more detail. }  of weight 1 and index $k$,
showing a much simpler modular behavior. 
It has been known \cite{Zwegers} that 
one can `absorb' this anomalous modularity into a suitable {\em non-holomorphic\/} correction term. 
Namely, we can define a function $\hf^{(k)}(\tau,z)$  of the form  
\begin{equation}
\hf^{(k)} (\tau,z) \equiv f^{(k)}(\tau,z)+ [\mbox{non-holomorphic correction}],
\label{completion}
\end{equation}
which becomes a non-holomorphic Jacobi form of weight 1 and index $k$. 
We shall call it the `modular completion'\footnote{In the mathematical language 
the modular completions correspond to the `harmonic Maass forms' associated to the mock modular forms. 
}
of $f^{(k)}(\tau,z)$ and will yield the precise definition of the correction term later.


So, what is the physical interpretation of the modular completion? 
When making the interpretation as characters of SCA, \eqn{completion} corresponds to the next schematic relation 
\begin{align}
\hchi_{\msc{dis}}(*;\tau,z) = \chi_{\msc{dis}}(*;\, \tau,z) + \sum\, [\mbox{{\em non-hol.}  coefficient}] \, \chi_{\msc{cont}}(*; \, \tau,z),
\label{completion 2}
\end{align}
where `$\chi_{\msc{dis}}$' denotes the character of discrete (or degenerate) representation while  `$\chi_{\msc{cont}}$' does 
the continuous (or non-degenerate) one. 
The second term of R.H.S would suggest a kind of mixing  of the discrete and continuous spectra, but 
the precise interpretation seems to be difficult due to the non-holomorphicity of the coefficients.  

A good laboratory to study this feature would be the  $SL(2,\br)/U(1)$-supercoset model,
which is the supersymmetric extension of 
the two-dimensional Euclidean black-hole \cite{2DBH}.
Among other things, it has been shown in \cite{ES-NH} that the modular completions are naturally obtained by evaluating 
the torus partition function  {\em by path-integration}.
This is remarkable in the sense that it has revealed a physical origin of the modular completion of mock modular form. 
Indeed, the modular transformations are identified as nothing but the coordinate transformations on the Euclidean torus of world-sheet
in any conformal field theories, and thus the good modularity {\em has to\/} be achieved as long as working with the path-integral. 
On the other hand, 
the `holomorphic factorization' such as 
$$
Z(\tau)\sim \sum_{j, \tj} \, N_{j,\tj}\, \chi_j(\tau) \, \overline{\chi_{\tj}(\tau)}, 
$$
is violated in this case, 
although it is naively expected in the `Hamiltonian formalism' or the representation theory of SCA;
$$
Z(\tau) = \tr \left[e^{-\beta H }\right].
$$
This incompatibility between the good modularity and holomorphicity
would originate from the existence of gapless continuous spectrum of non-BPS states 
causing the IR-divergence in this system.

We can also derive the modular completions of discrete characters by computing the elliptic genus in a simpler manner, 
as we will demonstrate in the next section. 
Then, the elliptic genus gets non-holomorphic although the expected modularity is gained. 
One may summarize this feature as follows;
{\em the Hamiltonian formalism respects the holomorphicity but produces the `modular anomaly',
whereas the path-integration leads to the good modularity but yields the `holomorphic anomaly'.}
Closely related studies are also given in \cite{Troost,AT,AT2,Murthy,AT3}.


~


We summarize main contents of this paper as follows;

~

\noindent
{\bf 1. }
In section 2, we evaluate the torus partition function and 
elliptic genus of the $SL(2,\br)/U(1)$-supercoset theory
based on the path-integration.
This model is a simple example of supercoset theories with $\cN=2$ superconformal symmetry \cite{KS}.
Nevertheless, it is fairly non-trivial to work with the partition function or the elliptic genus   
due to its non-compact nature or the non-rationality, and we face at the holomorphic anomaly noted above. 
We especially make use of the `spectral flow expansion' in the computation  of elliptic genus 
according to \cite{orb-ncpart,ES-nhP}, and derive simple formulas of the modular completions, 
which make their modular behavior manifest.


~


\noindent
{\bf 2. }
In section 3, 
as an application of our analyses on the $SL(2,\br)/U(1)$-theory, 
we study the elliptic genus of the ALE-spaces\footnote
   {The ALE-space is a four dimensional non-compact hyperK\"{a}hler manifold  in which the simple singularity (so-called  `ADE-type singularity') is  
resolved. Needless to say, $\mbox{ALE}(A_1)$ is identified with the famous Eguchi-Hanson solution of gravitational instanton \cite{EguchiHanson}. 
}
 realized as 
the non-compact Gepner-like orbifolds according to \cite{OV,GKP};
$$
\mbox{type II}/\mbox{ALE($G$)} \cong 
\left. SU(2)/U(1) \otimes SL(2)/U(1) \right|_{U(1)\msc{-charge} \in \bz},
$$
where the type of singularity $G$ for the ALE-space is naturally encoded into the data of modular invariance in the R.H.S. 
We especially review our work on this subject presented in
\cite{ES-N=4L}, motivated by the recent progress in studies of new types of moonshine phenomena 
\cite{EOT,umbral1,umbral2,CH,HM,HMN}.

We discuss a kind of duality between two $\cN=4$ superconformal systems of different central charges; 
one is the world-sheet CFT describing the ALE-background 
(or so-called the CHS-system  describing NS5-branes \cite{CHS}),
and the other would be identified as the $\cN=4$ supersymmetric extension of Liouville theory. 
From the view points of $\mbox{AdS}_3/\mbox{CFT}_2$-duality in the NS5-NS1 system studied in \cite{SW}, 
the latter should be identified as the `space-time CFT' generically possessing a  central charge proportional to 
the brane charges. 
Thus, this could offer a novel duality picture of moonshine phenomena; one is that of the world-sheet, and the other is of the space-time. 



~

~


\section{Torus Partition function and  Elliptic Genus of $SL(2)/U(1)$-Supercoset Model}

~

In this section, we review the studies on the torus partition function and elliptic genus of 
the $SL(2)/U(1)$-supercoset model presented mainly in \cite{ES-NH,orb-ncpart,ES-nhP}, 
which focus on the physical origin 
of the mock modular forms and  their modular completions.


~

\subsection{Torus Partition Function}


We start with demonstrating the computation of torus partition function of the $SL(2)/U(1)$-supercoset model
by performing the path-integration according to \cite{ES-NH}.
We shall incorporate the general twist angles $z, \bar{z} \in \bc$
(in the $\tR$-sector
) into the partition function for the reason that will get clear later. 
See also {\em e.g.} \cite{HPT,IPT,ES-BH} for the earlier works in which closely related 
analyses are presented. 

It is familiar that the supercoset theories are described by the SUSY gauged WZW models \cite{GawK,Schnitzer,KarS}.
We set the level of $SL(2, \br)$ super-WZW model to be a real positive number $k$ (the level of bosonic part
is $k+2$), which means that the central charge of this superconformal system is
\begin{equation}
\hc \left(\equiv \frac{c}{3} \right) = 1 + \frac{2}{k}.
\label{chat}
\end{equation}
The world-sheet action of relevant SUSY gauged WZW model for $SL(2,\br)/U(1)$
in our convention 
is written as $(\kappa\equiv k+2)$
\begin{eqnarray}
 S(g,A,\psi^{\pm}, \tpsi^{\pm}) &: =& \kappa S_{\msc{gWZW}}(g,A)
 + S_{\psi}(\psi^{\pm}, \tpsi^{\pm}, A), 
\label{total action}\\
\kappa S_{\msc{gWZW}} (g,A) &: =& \kappa S^{SL(2,\bsr)}_{\msc{WZW}} (g)
+ \frac{\kappa}{\pi}\int_{\Sigma}d^2v\, 
\left\{ A_{\bar{v}} \tr \left(\frac{\sigma_2}{2} \partial_v g g^{-1}\right)
\pm
\tr \left(\frac{\sigma_2}{2}g^{-1}\partial_{\bar{v}}g\right)A_v  
\right. \nn
&& \hspace{3cm} \left. 
\pm \tr\left(\frac{\sigma_2}{2}g\frac{\sigma_2}{2} g^{-1} \right)
   A_{\bar{v}}A_v 
+ \frac{1}{2} A_{\bar{v}}A_v  \right\} , 
\label{gWZW action} \\
S^{SL(2,\bsr)}_{\msc{WZW}} (g) &: =& -\frac{1}{8\pi} \int_{\Sigma} d^2v\,
\tr \left(\partial_{\al}g^{-1}\partial_{\al}g\right) +
\frac{i}{12\pi} \int_B \,\tr\left((g^{-1}dg)^3\right) ,
\label{SL(2) WZW action} 
\\
 S_{\psi}(\psi^{\pm}, \tpsi^{\pm}, A)&: =& 
\frac{1}{2\pi}\int d^2v\, \left\{
\psi^+(\partial_{\bar{v}}+A_{\bar{v}}) \psi^-
+\psi^-(\partial_{\bar{v}}-A_{\bar{v}}) \psi^+ \right. \nn 
&& \hspace{3cm} \left. +\tpsi^+(\partial_{v}\pm A_{v}) \tpsi^-
+\tpsi^-(\partial_{v}\mp A_{v}) \tpsi^+
\right\} .
\label{fermion action}
\end{eqnarray}
In \eqn{gWZW action} and \eqn{fermion action}, the $+$ sign/ $-$ sign  is chosen for the axial-like/vector-like  
gauged WZW model, which we shall denote as 
$S^{(A)}_{\msc{gWZW}}$/$S^{(V)}_{\msc{gWZW}}$ \,
($S^{(A)}_{\psi}$/$S^{(V)}_{\psi}$ )
from now on. 
The $U(1)$ chiral gauge transformation is defined 
by 
\begin{eqnarray}
&& g~ \longmapsto ~  \Om_L \, g \,  \Om_R^{\ep} , 
\nn
&&
A_{\bar{v}}~ \longmapsto ~ 
A_{\bar{v}} - \Om_L^{-1} \del_{\bar{v}} \Om_L, ~~~
A_{v}~ \longmapsto ~ 
A_{v}- \Om_R^{-1} \del_{v} \Om_R,
\nn
&&  
\psi^{\pm} ~ \longmapsto ~ \Om_L^{\pm 1} \, \psi^{\pm},
~~~ 
\tpsi^{\pm} ~ \longmapsto ~ \Om_R^{\pm \ep}  \, \tpsi^{\pm},
\nn
&&
\hspace{4cm}
\Om_L(v,\bar{v}), ~ \Om_R(v,\bar{v}) \in e^{i\br \sigma_2}, 
\label{chiral gauge trsf}
\end{eqnarray}
where we set $\ep\equiv +1, -1$ for the axial, vector model, respectively.
The gauged WZW action $S^{(A)}_{\msc{gWZW}}$ /  $S^{(V)}_{\msc{gWZW}}$ 
is invariant under the axial/vector type gauge transformations that 
correspond to $\Om_L=\Om_R 
$ in \eqn{chiral gauge trsf}.
Both of the classical fermion actions $S^{(A)}_{\psi}$, $S^{(V)}_{\psi}$ 
\eqn{fermion action} are invariant under general chiral gauge transformations
$\Om_L$, $\Om_R$, and we assume the absence of chiral anomalies 
when $\Om_L=\Om_R$ holds
(in other words, the anomalies emerge for $\Om_L=\Om_R^{-1}$).
It is well-known that this model describes the (supersymmetric extension of)
string theory on 2D Euclidean 
black-hole, or the `cigar geometry' \cite{2DBH}, and also known to be 
equivalent (mirror) to the $\cN=2$ Liouville theory \cite{HK}. 

It will be convenient to introduce alternative notations of gauged WZW actions; 
\begin{eqnarray}
&& S_{\msc{gWZW}}^{(A)} (g, h_L, h_R) :=  S^{SL(2,\bsr)}_{\msc{WZW}}(h_L g h_R) 
- S^{SL(2,\bsr)}_{\msc{WZW}}(h_L h_R^{-1}),
\label{gWZW A}
\\
&& S_{\msc{gWZW}}^{(V)} (g, h_L, h_R) :=  S^{SL(2,\bsr)}_{\msc{WZW}}(h_L g h_R) 
- S^{SL(2,\bsr)}_{\msc{WZW}}(h_L h_R).
\label{gWZW V}
\end{eqnarray}
They are equivalent with \eqn{gWZW action} under the identification of 
gauge field;
\begin{equation}
A_{\bar{v}}\frac{\sigma_2}{2} = \partial_{\bar{v}} h_L h_L^{-1}, 
\hspace{1cm}
A_v \frac{\sigma_2}{2} = \ep \, \partial_v h_R h_R^{-1},
\label{identification A} 
\end{equation} 
where we set $\ep =+1  , \, (-1)$ for the axial (vector) model as before.

Now,  transforming the world-sheet coordinates as $v=e^{iw}$, $\bar{v}=e^{-i\bar{w}}$,  
we define the world-sheet torus $\Sigma$ 
by the identifications  
$(w,\bar{w}) \sim (w+2\pi,\bar{w}+2\pi)
\sim(w+2\pi \tau, \bar{w}+2\pi \bar{\tau})$ ($\tau\equiv
\tau_1+i\tau_2$, $\tau_2>0$).
Including the `angle parameter' $z$ coupled with the $U(1)_R$-charge in $\cN=2$ SCA,
the torus partition function is written as 
\begin{eqnarray}
&& Z(\tau,  z) 
= \int_{\Sigma}\frac{d^2u}{\tau_2} \, 
\int \cD\lb g, A[u], \psi^{\pm}, \tpsi^{\pm}\rb\, 
\nn
&& \hspace{2cm} \times 
\exp \left[-\kappa S^{(A)}_{\msc{gWZW}}\left(g,A[u+\frac{2}{k}z]\right) - 
S^{(A)}_{\psi}\left(\psi^{\pm},\tpsi^{\pm}, A[u+\frac{k+2}{k} z ]\right)\right],
\label{part fn A 0}
\end{eqnarray}
where $\frac{d^2u}{\tau_2}$
is the modular invariant measure 
of modulus parameter $u$, and we work in  the $\tR$-sector for world-sheet fermions. 
%
We here adopted   
the notation `$A[u]$' for the gauge field 
$A\equiv \left(A_w dw + A_{\bar{w}} d\bar{w}\right) \frac{\sigma_2}{2}$ 
to emphasize the modulus dependence, 
explicitly written as 
\begin{equation}
A[u]_{\bar{w}}= \partial_{\bar{w}} X  + i \partial_{\bar{w}} Y 
- \frac{u}{2\tau_2}, 
~~~  A[u]_{w}=  \partial_{w} X  - i \partial_{w} Y
- \frac{\bar{u}}{2\tau_2}
\label{Au} 
\end{equation}
where $X$, $Y$ are real scalar fields 
parameterizing the chiral gauge transformations;
\begin{equation}
\Om_L = e^{(X+iY) \frac{\sigma_2}{2}}, ~~~ 
\Om_R \left( \equiv \Om_L^{\dag}\right) 
= e^{(X-iY) \frac{\sigma_2}{2}}.
\label{Om X Y}
\end{equation}
Note that the modulus parameter $u$ is normalized
so that it correctly couples with the zero-modes of $U(1)$-currents
$J^3 \equiv j^3 + \psi^+\psi^-$, $\tJ^3 \equiv \tj^3 + \tpsi^+ \tpsi^-$ 
which should be gauged (where $j^3$, $\tj^3$ are the bosonic parts).
On the other hand, 
the complex parameter $z$ precisely 
couples with the zero-modes of 
$\cN=2$ $U(1)_R$-currents $J$, $\tJ$ in the Kazama-Suzuki model \cite{KS};
\begin{equation}
J = 
\psi^+ \psi^- + \frac{2}{k} 
J^3
\equiv \frac{k+2}{k} \psi^+  \psi^- + \frac{2}{k} j^3,
\label{N=2 U(1)}
\end{equation}
($\tJ$ is defined in the same way.)


We can evaluate this path-integration 
by separating
the degrees of freedom of chiral gauge transformations $X$, $Y$
according to the standard quantization procedures 
of gauged WZW models \cite{KarS,GawK,Schnitzer}.
We have to path-integrate the compact boson $Y$ explicitly, while the
non-compact boson $X$ is decoupled as a gauge volume. 
By using the definitions of gauged WZW actions 
\eqn{gWZW A}, \eqn{gWZW V} 
and a suitable change of integration variables, 
we obtain
\begin{eqnarray}
 Z(\tau,  z) 
&=&  \int_{\Sigma}\frac{d^2u}{\tau_2} \, 
\int \cD\lb g, Y , \psi^{\pm}, \tpsi^{\pm}, b, \widetilde{b}, c, \widetilde{c} \rb\, 
\nn
&&  \times 
\exp \left[-\kappa S^{(V)}_{\msc{gWZW}}
\left(g,h^{u+\frac{2}{k}z}, 
\left(h^{u+\frac{2}{k}z}\right)^{\dag} \right) 
+
\kappa S^{(A)}_{\msc{gWZW}}
\left(e^{iY \sigma_2},
h^{u+\frac{2}{k}z}, 
h^{u+\frac{2}{k}z}
\right)
\right]
\nn
&& \times
\exp \left[
 -2 S^{(A)}_{\msc{gWZW}}
\left(e^{iY \sigma_2},h^{u+\frac{k+2}{k}z}, 
h^{u+\frac{k+2}{k}z}\right)
- S^{(A)}_{\psi}\left(\psi^{\pm},\tpsi^{\pm}, a[u+\frac{k+2}{k} z ]\right)
\right]
\nn
&& \times 
\exp \left[
- S_{\msc{gh}}(b,\widetilde{b}, c,  \widetilde{c})\right],
\label{part fn A 1}
\end{eqnarray}
where the ghost variables have been introduced to rewrite the Jacobian factor. 
It is most non-trivial to evaluate the path-integration of the compact boson $Y$.
Its world-sheet action is evaluated as  
\begin{eqnarray}
S^{(A)}_Y(Y,u, z)& \equiv & 
- \kappa S^{(A)}_{\msc{gWZW}}
\left(e^{iY \sigma_2},
h^{u+\frac{2}{k}z}, 
\left(h^{u+\frac{2}{k}z}\right)^{\dag}
\right)
+ 2 S^{(A)}_{\msc{gWZW}}
\left(e^{iY \sigma_2},h^{u+\frac{k+2}{k}z}, 
\left(h^{u+\frac{k+2}{k}z}\right)^{\dag}\right)
\nn
& = & \frac{k}{\pi} \int_{\Sigma} d^2 v\, \del_{\bar{w}} Y^u \del_w Y^u 
- \frac{2\pi}{\tau_2} \hc \left|z\right|^2,
\label{S Y A}
\end{eqnarray}
where we introduced the twisted scalar filed 
$Y^u \equiv Y + \frac{1}{\tau_2} \mbox{Im}\, (w \bar{u})$ 
that satisfies the following boundary condition
($u\equiv s_1 \tau + s_2$, $0\leq s_1, s_2 \leq 1$);
\begin{eqnarray}
&&Y^u(w+2\pi,\bar{w}+2\pi)=Y^u(w,\bar{w})- 2\pi (m_1+s_1), \nn
&&Y^u(w+2\pi\tau,\bar{w}+2\pi\bar{\tau})=Y^u(w,\bar{w})+ 2\pi (m_2+s_2), 
~~~(m_1, m_2 \in \bz).
\label{Yu bc}
\end{eqnarray}
Note that the linear couplings between the $U(1)$-currents $i\del_w Y$, $i\del_{\bar{w}} Y$
and the $\cN=2$ moduli $z$, $\bar{z}$ are precisely canceled out.
This should be the case, since the $\cN=2$ $U(1)$-current $J$ possesses no contribution from the compact boson $Y$, 
which should be gauged away.

Because of the boundary condition \eqn{Yu bc},
the zero-mode integral yields the summation over winding sectors weighted by 
the factor $e^{-\frac{\pi k}{\tau_2} \left|m_1\tau+m_2+ u\right|^2}
\equiv e^{-\frac{\pi k}{\tau_2} \left|(m_1+s_1)\tau+(m_2+s_2)\right|^2}$ determined 
by the `instanton action'.
After all, we achieve the next formula of partition function \cite{ES-NH};
\begin{eqnarray}
Z(\tau,z) &=& 
\cN \, e^{\frac{2\pi}{\tau_2} \left( \hc \left|z\right|^2 - \frac{k+4}{k} z_2^2\right)}\,
\sum_{m_1,m_2\in \bz}\, \int_{\Sigma} \frac{d^2 u}{\tau_2} \, 
\left|
\frac{\th_1\left(\tau,u+\frac{k+2}{k}z\right)}{\th_1\left(\tau, u+\frac{2}{k}z\right)}
\right|^2 \,
e^{-4\pi \frac{u_2z_2}{\tau_2}} \, e^{-\frac{\pi k}{\tau_2} \left|m_1\tau+m_2+ u\right|^2}
\nn 
&\equiv & \cN \, e^{\frac{2\pi}{\tau_2} \left( \hc \left|z\right|^2 - \frac{k+4}{k} z_2^2\right)}\,
\int_{\bc} \frac{d^2 u}{\tau_2} \, 
\left|
\frac{\th_1\left(\tau,u+\frac{k+2}{k}z\right)}{\th_1\left(\tau, u+\frac{2}{k}z\right)}
\right|^2 \,
e^{-4\pi \frac{u_2z_2}{\tau_2}} \, e^{-\frac{\pi k}{\tau_2} \left|u\right|^2},
\label{part fn A}
\end{eqnarray}
where $\cN$ is a normalization constant. 
This is identified as the Euclidean cigar model whose asymptotic circle has the radius $\sqrt{\al' k}$.

The $u$-integration leads to the obvious IR-divergence
that originates from the non-compactness of cigar geometry.
Therefore,  we need to introduce a regularization scheme. 
We adopt the following regularized partition function
\footnote
{We shall denote the regularized partition function as `$Z_{\msc{reg}}(\tau,z, \bar{z}; \ep)$'
rather than `$Z_{\msc{reg}}(\tau,z ; \ep)$' although $\bar{z}$ is just the complex conjugate of $z$ 
in \eqn{Z reg}. This is because we will later treat $z$ and $\bar{z}$ as two {\em independent\/} complex
variables to derive the elliptic genus, while $\bar{\tau}$ 
is always the complex conjugate of $\tau$ 
in this section.
}
($u\equiv u_1+i u_2\equiv s_1 \tau+ s_2, ~ z\equiv z_1 +i z_2,~~ u_1, u_2, s_1,s_2,
 z_1, z_2 \in \br$) 
\begin{eqnarray}
Z_{\msc{reg}}(\tau,z, \bar{z}; \ep) &=&  
\cN \, e^{\frac{2\pi}{\tau_2} \left( \hc \left|z\right|^2 - \frac{k+4}{k} z_2^2\right)}\,
\int_{\bc} \frac{d^2 u}{\tau_2}\, 
\sigma_{\ep}(\tau, u,z, \bar{z})\,
\left|
\frac{\th_1\left(\tau, u + \frac{k+2}{k}z\right)}
{\th_1\left(\tau, u + \frac{2}{k} z \right)}\right|^2 
\, e^{- 4\pi  z_2 \frac{u_2}{\tau_2}}\,
e^{-\frac{\pi k}{\tau_2}\left| u \right|^2},
\nn
&&
\label{Z reg}
\end{eqnarray}
where we introduced a regularization factor 
$\sigma_{\ep}(\tau, u,z, \bar{z})$ 
($\ep >0$)
defined as
\begin{eqnarray}
\sigma_{\ep}(\tau, u,z,\bar{z}) := \prod_{m_1,m_2 \in \bz} \, \left[ 1 - 
 e^{-\frac{1}{\ep \tau_2} 
\left\{ (s_1+m_1) \tau+ (s_2+m_2) + \frac{2z}{k} \right\}
\left\{ (s_1+m_1) \bar{\tau}+ (s_2+m_2) + \frac{2\bar{z}}{k} \right\}
}\right].
\label{def sigma}
\end{eqnarray}
All the singularities of the integrand located at $u + \frac{2}{k} z \in \bz \tau + \bz$
that originate from the $\th_1$-factor 
are removed by inserting  \eqn{def sigma}
and the $u$-integral converges as long as $\ep>0$.
We will later regard  $\sigma_{\ep}(\tau, u, z, \bar{z})$ as a holomorphic function
with respect to {\em complex\/} variables $s_1, s_2$.

Now, the $u$-integral in  
\eqn{Z reg} is finite  as long as $\ep >0$,
and because of the modular invariance 
of $\sigma_{\ep} (\tau, u,z,\bar{z})$, 
$Z_{\msc{reg}}(\tau,z, \bar{z}; \ep)$
is defined 
so as to preserve the expected modular invariance;
\begin{equation}
Z_{\msc{reg}}(\tau+1,z, \bar{z}; \ep) = Z_{\msc{reg}}(\tau,z, \bar{z}; \ep) ,
\hspace{1cm} 
Z_{\msc{reg}}\left(-\frac{1}{\tau}, \frac{z}{\tau}, \frac{\bar{z}}{\bar{\tau}}; \ep\right) 
= Z_{\msc{reg}}(\tau,z, \bar{z}; \ep). 
\end{equation}
The partition function $Z_{\msc{reg}}(\tau,z, \bar{z}; \ep) $
logarithmically diverges in  the limit $\ep\, \rightarrow\, +0$, 
and the divergent part is identified with the contributions from  
the strings propagating in the asymptotic cylindrical region of the cigar geometry.
It implies that the characteristic behavior around $\ep \, \rightarrow \, +0$ is given by 
\begin{eqnarray}
&& Z_{\msc{reg}}(\tau, z, \bar{z}; \ep) = \cC 
| \ln \ep |
e^{2\pi \frac{\hc}{\tau_2} z_1^2}\,
\left|\frac{\th_1(\tau,z)}{\eta(\tau)^3}\right|^2\,
\nn
&& \hspace{3cm}
\times
\sum_{n,w\in \bz}\,\int _0^{\infty}dp\, 
q^{\frac{p^2}{2} + \frac{1}{4}\left(\frac{n}{\sqrt{k}} + \sqrt{k} w\right)^2}
\overline{
q^{\frac{p^2}{2} + \frac{1}{4}\left(\frac{n}{\sqrt{k}} - \sqrt{k} w\right)^2}
}\, y^{\frac{n+kw}{k}} \overline{y^{-\frac{n-kw}{k}}}
\nn
&& \hspace{6cm}
+ Z_{\msc{finite}}(\tau, z, \bar{z}) + O(\ep, \ep\ln \ep),
\label{asp part}
\end{eqnarray}
where $\cC$ is some positive constant independent of $\ep$. 
The leading part proportional to the volume factor
$\left| \ln \ep \right|$
(the `asymptotic part')
is  expressible in terms of the extended continuous characters 
\cite{ES-L,ES-BH} as in the free theory 
when the level $k$ is rational. 
We here denote the term of order $O(\ep^0)$ as  
$Z_{\msc{finite}}(\tau, z, \bar{z})$, 
which is still finite after taking the $\ep\, \rightarrow \, +0$ limit and is 
modular invariant. This part can be directly extracted from 
$Z_{\msc{reg}}(\tau,z,\bar{z} ; \ep)$ as given in \cite{ncpart-NENS5};
\begin{equation}
Z_{\msc{finite}}(\tau, z, \bar{z}) = \lim_{\ep\,\rightarrow\, +0}\,
\left[1- \ep \ln \ep \frac{\del}{\del \ep} \right] Z_{\msc{reg}}(\tau,z,\bar{z} ; \ep).
\label{finite part formula}
\end{equation}
We emphasize that $Z_{\msc{finite}}(\tau, z, \bar{z})$ is uniquely determined 
irrespective of the adopted regularization scheme, even though the overall constant $\cC$ in the asymptotic part 
as well as the correction terms of $O(\ep, \ep \ln \ep)$
will depend on the method of regularization.


~

\subsection{Elliptic Genus}

~

We next consider the elliptic geuns \cite{Witten-E} of the $SL(2,\br)/U(1)$-theory.
This is obtained by formally setting $\bar{z}=0$ while fixing $z$ 
at a general complex value in $Z_{\msc{finite}}(\tau, z, \bar{z})$
\eqn{finite part formula}. 
We also need to divide the partition function by the factor
\begin{equation}
e^{2\pi \frac{\hc}{\tau_2} \left( |z|^2 - z_2^2\right)  } \equiv e^{2\pi  \frac{\hc}{\tau_2} \left(\frac{z+\bar{z}}{2}\right)^2} 
\sim e^{\frac{\pi}{\tau_2} \frac{\hc}{2}z^2},
\label{anomaly factor 1}
\end{equation}
in order to include the anomaly factor correctly.
In fact, we can uniquely determine this factor by requiring the following conditions due to the analysis given in \cite{ES-NH};
\begin{itemize}
\item The elliptic genus should have the correct modular property;
\begin{eqnarray}
&& \cZ(\tau+1,z) = \cZ(\tau,z), \hspace{1cm}
\cZ\left(-\frac{1}{\tau}, \frac{z}{\tau} \right) 
= e^{i\pi \frac{\hc}{\tau} z^2}\, 
\cZ(\tau,z),
\label{modular EG}
\end{eqnarray}

\item The elliptic genus should be expanded by the discrete characters \eqn{chd} 
around  $\tau \sim i\infty$ as 
\begin{eqnarray}
\cZ(\tau,z) & = &  \sum\, \chd(*,0;\tau,z) + [\mbox{subleading terms}],
\label{IR part EG}
\end{eqnarray}
{\em with no extra overall 
factor\/},  
where the `subleading terms' include 
spectral flow sectors as well as non-holomorphic corrections. 

\end{itemize}
In this way, 
the elliptic genus is written as 
\begin{eqnarray}
\cZ(\tau,z) & = & e^{-\frac{\pi}{\tau_2} \frac{\hc}{2}z^2} \,  
Z_{\msc{finite}}(\tau,z,\bar{z}=0)
\nn
& \equiv & \lim_{\ep \rightarrow +0}\, 
e^{-\frac{\pi}{\tau_2} \frac{\hc}{2}z^2} \,  
Z_{\msc{reg}}(\tau,z,\bar{z}=0 ; \ep).
\label{EG path int 0}
\end{eqnarray}
The equality of second line is owing to the simple fact that 
the asymptotic term in \eqn{asp part} drops off when setting $\bar{z}=0$.
We thus obtain the path-integral expression of elliptic genus as 
\begin{eqnarray}
\hspace{-5mm}
\cZ(\tau,z) &=& 
\lim_{\ep \rightarrow +0}\, 
k e^{\frac{\pi z^2}{k\tau_2}} \, \int_{\bc} \frac{d^2 u}{\tau_2}\, 
\sigma_{\ep}(\tau, u,z,0)\,
\frac{\th_1\left(\tau, u + \frac{k+2}{k}z\right)}
{\th_1\left(\tau, u + \frac{2}{k} z \right)} \, 
e^{2\pi i z \frac{u_2}{\tau_2}}\,
e^{-\frac{\pi k}{\tau_2}\left| u \right|^2}.
\label{EG path int}
\end{eqnarray}
The $u$-integral is not  easy to perform since 
the Gaussian factor 
$e^{-\frac{\pi k}{\tau_2} \left| u \right|^2}$ 
breaks the periodicity of the integrand.
If this factor was absent, the relevant integral would reduce to 
a simple period integral over a torus $\Sigma \equiv \bc/\La$, ($\La \equiv \bz\tau+\bz$).
One way to avoid this complication is given by using 
the following identity \cite{orb-ncpart,ES-nhP}, which we shall call the `spectral flow expansion';
\begin{equation}
\cZ(\tau,z) = \sum_{\la \equiv n_1\tau + n_2 \in \La}\,
(-1)^{n_1+n_2 + n_1 n_2}\,  
s^{(\frac{\hc}{2})}_{\la} \cdot \cZ^{(\infty)}(\tau,z).
\label{Z sflow formula}
\end{equation}
where $s^{(\kappa)}_{\la}$ denotes the spectral flow 
operator defined by
\begin{eqnarray}
s^{(\kappa)}_{\la} \cdot f(\tau,z) &:= &
q^{\kappa \al^2} y^{2\kappa \al} e^{2\pi i \kappa \al \beta}\,
f(\tau,z+\la)
\nn
& \equiv & e^{2\pi i \frac{\kappa}{\tau_2} \la_2 \left(\la + 2z\right) }\,
f(\tau, z+\la),
\nn
&& \hspace{1cm} 
(\la \equiv \al \tau +\beta\equiv \lambda_1+i\lambda_2, ~ \al, \beta, \lambda_1,\lambda_2 \in \br).
\label{sflow op}
\end{eqnarray}
Recall that the elliptic genus of a complex $D$-dimensional manifold 
is a Jacobi form with index ${D\over 2}$ \cite{Eich-Zagier}. Here $\hat{c}$ (\ref{chat}) is 
the effective dimension of a target manifold described by a superconformal field theory with a central charge $c$. 
Thus the suffix ${\hat{c} \over 2}$ of the flow operator $s_{\lambda}^{({\hat{c}\over 2})}$ denotes the index of the 
elliptic genus $\cZ(\tau,z)$ describing the cigar geometry.


On the other hand, $\cZ^{(\infty)}(\tau,z)$ is defined as 
the elliptic genus of `$\bz_{\infty}$-orbifold' of the cigar model, 
or equivalently the universal cover of trumpet in the $T$-dual picture. 
Namely, 
\begin{eqnarray}
\hspace{-1cm}
\cZ^{(\infty)}(\tau,z) 
&: =& 
\lim_{\ep \rightarrow +0}\, 
k e^{\frac{\pi z^2}{k\tau_2}} \, 
\int_{\Sigma} \frac{d^2 \om}{\tau_2} \,
\int_{\bc} \frac{d^2 u}{\tau_2}\,
\sigma_{\ep}(\tau, u ,z,0)\,
\frac{\th_1\left(\tau, \mu + \frac{k+2}{k}z\right)}
{\th_1\left(\tau, \mu  + \frac{2}{k} z \right)} 
\, e^{2\pi i z \frac{u_2}{\tau_2}}\,
e^{-\frac{\pi k}{\tau_2}\left| u+ \om \right|^2}
\nn
&=& 
\lim_{\ep \rightarrow +0}\, 
 e^{\frac{\pi z^2}{k\tau_2}} \, 
\int_{\Sigma} \frac{d^2 u}{\tau_2}\, 
\sigma_{\ep}(\tau, u ,z,0)\,
\frac{\th_1\left(\tau, u + \frac{k+2}{k}z\right)}
{\th_1\left(\tau, u  + \frac{2}{k} z \right)} 
\, e^{2\pi i z \frac{u_2}{\tau_2}}.
\label{eval EG infty}
\end{eqnarray}
In the second line, 
due to the periodicity of the integrand
under $u\, \rightarrow \, u + \nu$, $\nu \in \La$
 except for the factor 
$
e^{-\frac{\pi k}{\tau_2}\left| u+ \om \right|^2}
$, 
we made use of an obvious relation 
$
\int_{\Sigma} \frac{d^2 \om}{\tau_2} \, 
\int_{\bc} \frac{d^2 u}{\tau_2}\, \left[\cdots \right]
= \int_{\bc} \frac{d^2 \om}{\tau_2} \, 
\int_{\Sigma} \frac{d^2 u}{\tau_2}\, \left[\cdots \right],
$
and  carried out the Gaussian integral over $\om$.

Computation is now easy.
Since the integrand of \eqn{eval EG infty}  
is holomorphic and periodic with respect to the integration
variables $s_1$, $s_2$ with 
$u \equiv s_1\tau+s_2$, 
one may regard it as a double period integral.
Thus,  by deforming the integration contour as 
\begin{equation}
s_1 \in [0,1] + i\frac{z}{k\tau_2}, \hspace{1cm}
s_2 \in [0,1] - i\frac{z\bar{\tau}}{k\tau_2}, 
\label{contour deform 1}
\end{equation}
we can directly evaluate \eqn{eval EG infty}  
with the helps of the identity \eqn{theta id} as 
\begin{eqnarray}
\cZ^{(\infty)}(\tau,z) & = & \lim_{\ep\, \rightarrow \, +0} 
e^{- \frac{\pi z^2}{k\tau_2}}\, 
\int_{\Sigma} \frac{d^2 u}{\tau_2}\, 
\sigma_{\ep}(\tau, u,0, 0)
\frac{\th_1\left(\tau, u + z\right)}
{\th_1\left(\tau, u  \right)} \, 
\, e^{2\pi i z \frac{u_2}{\tau_2}}
\nn
&=& 
e^{- \frac{\pi z^2}{k\tau_2}}\, 
\int_{\Sigma} \frac{d^2 u}{\tau_2}\, 
\frac{\th_1\left(\tau, u + z\right)}
{\th_1\left(\tau, u  \right)} \, 
\, e^{2\pi i z \frac{u_2}{\tau_2}}
\nn
&=& e^{- \frac{\pi z^2}{k\tau_2}}\, 
\frac{\th_1(\tau,z)}{i\eta(\tau)^3}\, \sum_{n\in \bz}\, \int_{\Sigma} \frac{d^2 u}{\tau_2}\, 
\frac{e^{2\pi i n u }}{1-yq^n}\, e^{2\pi i z \frac{ u_2}{\tau_2}}
\nn
&=& e^{- \frac{ \pi z^2}{k\tau_2}}\,  
\frac{\th_1(\tau,z)}{2\pi z \eta(\tau)^3}.
\label{EG infty}
\end{eqnarray}
In the second line, we have used  
$$
\sigma_{\ep}(\tau, u,0,0) = 
 \prod_{m_1,m_2 \in \bz}\, 
\left[ 1- e^{- \frac{1}{\ep \tau_2} \left|u + m_1 \tau+m_2\right|^2} \right],
$$
as well as the fact that the $u$-integral converges even in the absence of 
$\sigma$-factor, contrary to the previous evaluation of torus partition function.  

Substituting \eqn{EG infty} back into the formula of spectral flow expansion  
\eqn{Z sflow formula},  
we finally obtain the following very simple expression of elliptic genus of cigar model;
\begin{eqnarray}
\cZ(\tau,z) & = & 
\sum_{\la \equiv n_1\tau + n_2 \in \La}\,
(-1)^{n_1+n_2 + n_1 n_2}\,  
s^{(\frac{\hc}{2})}_{\la} \cdot
\left[
e^{- \frac{ \pi z^2}{k\tau_2}}\,  
\frac{\th_1(\tau,z)}{2\pi z \eta(\tau)^3}
\right]
\nn
& =& 
\frac{\th_1(\tau,z)}{2\pi  \eta(\tau)^3} 
\sum_{\la \in \La}\,
s^{(\frac{1}{k})}_{\la} \cdot
\left[
\frac{e^{- \frac{ \pi z^2}{k\tau_2}}}{z}
\right]
\nn
& = &  
\frac{\th_1(\tau,z)}{2\pi  \eta(\tau)^3} 
\sum_{\la \in \La}\, 
\frac{
e^{- \frac{\pi}{k\tau_2} \left[ z^2+  |\la|^2 + 2 \bar{\la} z \right] }
}
{z+\la}
\equiv 
\frac{\th_1(\tau,z)}{2\pi  \eta(\tau)^3} 
\sum_{\la \in \La}\, 
\frac{
\rho^{(1/k)}(\la,z)
}
{z+\la}.
\label{EG nh-E}
\end{eqnarray}
In the last line, we introduced the conventional symbol
\begin{equation}
\rho^{(\kappa)}(\la,z) := s^{(\kappa)}_{\la} \cdot 
e^{-\frac{\pi \kappa}{\tau_2} z^2} \equiv 
e^{- \frac{\pi \kappa}{\tau_2} \left[|\la|^2 + 2\bar{\la} z + z^2\right] }.
\label{def rho}
\end{equation}
More explicitly, \eqn{EG nh-E} is rewritten as 
\begin{eqnarray}
\cZ(\tau,z) = 
\frac{\th_1(\tau,z)}{2\pi  \eta(\tau)^3} 
\, 
\sum_{m, n\in \bz}\, q^{\frac{1}{k} m^2} y^{\frac{2}{k} m} e^{2\pi i \frac{mn}{k}}\, 
\frac{ e^{-\frac{\pi }{k \tau_2} (z+m\tau+n)^2}}{z+m\tau+n}.
\label{EG nh-E 2}
\end{eqnarray}

Here the double power series of $\la \in \La$
absolutely converges due to the Gaussian factor, 
and thus \eqn{EG nh-E} exhibits the good modular behavior.
One can easily confirm that the elliptic genus \eqn{EG nh-E} 
possesses the modular property as a weak Jacobi form with weight 0 and index $\hat{c}/2$ given in \eqn{modular EG},
and also 
\begin{eqnarray}
&& s^{(\frac{\hc}{2})}_{n_1\tau+n_2} \cdot \cZ(\tau,z) = (-1)^{n_1+n_2 + n_1 n_2}
 \cZ(\tau,z), \hspace{1cm}
 (\any n_1, n_2 \in N \bz),
 \label{sflow EG}
\end{eqnarray}
for the case of $k= N/K$, $N, K \in \bz_{>0}$,

~


\subsection{Relations to Mock Modular Forms}

It is amusing that the elliptic genus of $SL(2)/U(1)$-theory 
is written in terms of the modular completion 
of the mock modular form, which was first shown in  
\cite{Troost, ES-NH}. 
The relevant mock modular form\footnote
   {See {\em e.g.} \cite{DMZ} for the precise definitions and mathematically rigid terminologies 
for mock modular forms.} is called in many literature as 
the `Appell-Lerch sum' \cite{Zwegers,STT}, 
defined by 
\begin{equation}
f^{(k)}_u(\tau,z)
:=  \sum_{n\in \bsz} 
\frac{q^{ k n^2} y^{2 kn} }{1-y w^{-1} q^n} ,
\hspace{1cm} 
(q\equiv e^{2\pi i\tau}, ~ y\equiv e^{2\pi i z}, ~ w\equiv e^{2\pi i u} ),
\label{Appell u}
\end{equation}
and its `modular completion' is explicitly given as \cite{Zwegers};
\begin{equation}
\widehat{f}_u^{(k)}(\tau,z)
:=   
f^{(k)}_u(\tau,z)
- \frac{1}{2} \sum_{m\in \bz_{2k}} \, R_{m,k}(\tau, u) \Th{m}{k} (\tau, 2z), 
\hspace{1cm} (k\in \bz_{>0}), 
\label{hAppell}
\end{equation}
with\footnote
{
See Appendix A for the convention of error function $\erf(x)$, theta function $\Th{m}{k}(\tau,z)$ and Jacobi forms.
}
\begin{eqnarray}
&&R_{m,k}(\tau,u) := \sum_{\nu \in m+ 2k \bz} \,
\left[ \sgn(\nu+0) -  
\erf\left\{ \sqrt{\frac{\pi \tau_2}{k}} \left(\nu+ 2k \frac{u_2}{\tau_2} 
\right)\right\} \right]\,
w^{-\nu} q^{-\frac{\nu^2}{4k}},
\nn
&&
\hspace{9cm}
\left(\tau_2\equiv \Im \, \tau, ~ u_2 \equiv \Im \, u\right).
\label{Rmk}
\end{eqnarray}
Note that $R_{m,k}$ is non-holomorphic because of the inclusions of $\tau_2$ and $u_2$.
The holomorphic function $f^{(k)}_u(\tau,z)$ shows a complicated modular transformation law, 
which is often called the `mock modularity'.
In the physics context  this function with $u=0$ is closely related to 
the character formulas of massless representations of $\cN=4$ superconformal algebra (SCA) \cite{ET}
as well as the `extended characters' (spectral flow sums) of $\cN=2$ SCA \cite{Odake,ES-L}.
The modular property of $f^{(k)}(\tau,z) \equiv f^{(k)}_{u=0}(\tau,z)$ is summarized in \eqn{S Appell} and \eqn{T Appell}.

It is crucial that the anomalous modular transformation law of the `holomorphic part'  $f^{(k)}_u(\tau,z)$ 
is compensated by the second term in the R.H.S. of (\ref{hAppell}), 
and the `completed' one $\hf_u^{(k)}(\tau,z)$ behaves as a Jacobi form of weight $1$ and index $k$.
This means that the elliptic genus of $SL(2)/U(1)$-supercoset is described by  
a non-holomorphic generalization of a Jacobi form.


Let us clarify the precise relation between our computation of elliptic genus $\cZ(\tau,z)$ 
\eqn{EG nh-E} (or \eqn{EG nh-E 2}) and the modular completion $\hf^{(k)}(\tau,z) \equiv \hf^{(k)}_{u=0}(\tau,z)$.
In the case of $k=N/K$, ($N,K \in \bz_{>0}$), the formula \eqn{EG nh-E} is rewritten as 
\begin{eqnarray}
\cZ(\tau,z) & = & 
\frac{\th_1(\tau,z)}{i\eta(\tau)^3} \, \frac{1}{N} \sum_{a,b\in \bz_N}\,
s_{a\tau+b}^{(NK)} \cdot \hf^{(NK)}\left(\tau, \frac{z}{N}\right).
\label{EG cigar formula}
\end{eqnarray}
This expression has been originally derived in \cite{ES-NH}.
Equating this formula with \eqn{EG nh-E} in the special case of 
$N=1$, $K=1/k \in \bz_{>0}$,   
one can rewrite $\hf^{(K)}$
in a compact form;  
\begin{eqnarray}
\hf^{(K)}(\tau,z) & =& \frac{i}{2\pi} \,
\sum_{\nu \in \La} \, 
\frac{
\rho^{(K)}(\nu,z)}
{z+\nu}
\nn
&\equiv&
\frac{i}{2\pi} \sum_{m,n \in \bz }\, q^{K m^2 } y^{2K m} \frac{e^{-\frac{\pi K}{\tau_2} (z+m\tau+n)^2 }}{z+m\tau+n}.
\label{hAppell nh-E}
\end{eqnarray}
This is a fairly non-trivial identity and 
a direct proof of it has been presented 
in Appendix C of \cite{ES-nhP} with the `$u$-parameter' included.




We also point out the nice relation \cite{ES-NH,orb-ncpart};
\begin{eqnarray}
\cZ(\tau,z)  &=&  
\sum_{\stackrel{v, a \in \bz_N}{v+Ka \in N\bz}}\, 
\hchid^{(\stR)}(v,a;\tau,z) 
\nn
& \equiv &  \sum_{\stackrel{v, a \in \bz_N}{v+Ka \in N\bz}}\, 
\sum_{m\in a + N\bz}\, \hchd^{(\stR)}\left(\frac{v}{K}, m;\tau,z\right), 
\label{EG cigar formula 2}
\end{eqnarray}
where $\hchid^{(\stR)}$, $\hchd^{(\stR)}$ denote the modular completions of the discrete extended and irreducible characters  \eqn{hchid}, \eqn{hchd} 
of $\cN=2$ SCA with $\dsp \hc = 1 + \frac{2}{k} \equiv 1+ \frac{2K}{N}$.

In fact, this  would be expected from the viewpoints of representation theory of 
$\cN=2$ SCA
in the manner similar to the $SU(2)/U(1)$ Kazama-Suzuki model 
and the $\cN=2$ minimal characters \cite{Witten-E2}. 
However, it should be emphasized that $\cZ(\tau,z)$ is expanded by 
the {\em modular completions\/}, not by the characters themselves. 
We then likewise obtain 
\begin{eqnarray}
\hchid^{(\stR)}(v,a;\tau,z) & = & 
\frac{\th_1(\tau,z)}{2\pi \eta(\tau)^3}\, 
\sum_{b\in \bz_N}\, \sum_{\la\in a\tau+b + N\La} \,e^{-2\pi i \frac{b}{N}(v+Ka)}\,
\frac{\rho^{\left(1/k\right)}(\la,z)}{z+\la},
\label{hchid nh-E}
\\
\hchd^{(\stR)}(\nu,m;\tau,z)  &= &
\frac{\th_1(\tau,z)}{2\pi \eta(\tau)^3}\, 
\sum_{n\in \bz}\, e^{-2\pi i \frac{n}{k}(\nu+m)}\,
\frac{\rho^{(1/k)} (m\tau+n, z)}{z+m\tau+n}.
\label{hchd nh-E}
\end{eqnarray}


In the case of $k= N \in \bz_{>0}$ ($\hc = 1+ \frac{2}{N}$, in other words), 
we also introduce the function $\hF^{(N)}(v, a ; \tau,z)$ defined by
\begin{equation}
\hchid^{(\stR)}(v,a;\tau,z) \equiv \frac{\th_1(\tau,z)}{i\eta(\tau)^3} \, 
\hF^{(N)}(v, a ; \tau,z),
\label{def hF}
\end{equation}
for the convenience of arguments in the next section. 
This function is explicitly written as 
\begin{align}
\hF^{(N)}(v, a ; \tau,z)
& = \sum_{n\in a + \bz}\, 
\frac{\left(y q^n \right)^{\frac{v}{N}}}{1-yq^n}\, y^{\frac{2n}{N}} q^{\frac{n^2}{N}}
- \frac{1}{2} \sum_{j\in \bz_2}\, R_{v+N j, N} (\tau,0) \, 
\Th{v+Nj+2a}{N} \left(\tau, \frac{2z}{N}\right)
\nn
& \equiv \frac{i}{2\pi} \sum_{m \in a + N \bz}\, \sum_{n\in \bz}\, 
e^{- 2\pi i \frac{n}{N}(v+a)}\, \frac{\rho^{\left(\frac{1}{N}\right)}(m\tau+n, z)}{z+m\tau+n}.
\label{formula hF}
\end{align}
The second line is readily derived from 
\eqn{hchid nh-E}, and again manifestly exhibits the good modular property of it.


~


Some comments are in order:

\begin{description}
\item[1.]
In the above analysis we assumed the axial-like gauging, in other words, the cigar model. 
So, what happens if instead taking the vector-like  gauging?
As was examined in \cite{orb-ncpart}, in the case of $k=N/K$ ($N$ and $K$ are coprime positive integers), 
the vector-like $SL(2)/U(1)$-theory is identified with the $\bz_N$-orbifold of cigar model.
The elliptic genus of the vector-like model is computed as 
\begin{align}
\cZ_{\msc{vector}}(\tau,z)
& = 
\frac{\th_1(\tau,z)}{i\eta(\tau)^3} \, 
 \hf^{(NK)}\left(\tau, \frac{z}{N}\right)
\nn
& = \sum_{v\in \bz_N}\, \hchid^{(\stR)}(v,0;\tau,z).
\label{EG vector-like formula}
\end{align}
The expression in the first line has been originally  derived in \cite{Troost} for the case of $K=1$.

~


\item[2.]
In addition to the elliptic genera, 
the `finite part' of torus partition function $Z_{\msc{finite}}(\tau,z,\bar{z})$ 
\eqn{finite part formula} is also expressible in terms of the modular completions of extended characters $\hchid$.
For instance, in the cigar model we obtain \cite{ES-NH}
\begin{align}
& Z_{\msc{finite}}(\tau, z, \bar{z}) = e^{-2\pi \frac{\hc}{\tau_2}z_2^2}\,
\sum_{v=0}^{N-1}\,\sum_{\stackrel{a_L,a_R\in \bz_N}{v+K(a_L+a_R) \in N \bz}}\,
\hchid^{(\stR)}(v,a_L;\tau,z) \, \overline{\hchid^{(\stR)}(v, a_R;\tau,\bar{z})}.
\label{Z fin exp}
\end{align}
Again the modular completions $\hchid^{(\stR)}(v,a;\tau,z)$ play 
the similar role to those for the $\cN=2$ minimal characters in the $SU(2)/U(1)$ Kazama-Suzuki model.

~


\item[3.]
What is the role of the `$u$-parameter' of the function 
\eqn{hAppell} in the $SL(2)/U(1)$-gauged WZW model?
It is actually identified with
the {\em continuous\/} twist parameter 
of the general spin structure of the world-sheet fermions. 
More precisely, 
setting $u\equiv \al \tau+ \beta $, $(\any \al, \beta \in \br)$, 
we can extend the analyses given above to those with the world-sheet fermions of 
$\psi^{\pm} (w)$, $\tpsi^{\pm}(\bar{w})$ 
satisfy the twisted  boundary condition (with respect to the cylinder coordinate $w$, $\bar{w}$); 
\begin{eqnarray}
&& \psi^{\pm}(w+2\pi) = e^{\mp 2\pi i \al } \psi^{\pm}(w), \hspace{1cm} 
\psi^{\pm} ( w+ 2\pi \tau) = e^{\mp 2\pi i \beta} \psi(w),
\nn
&& \tpsi^{\pm} (\bar{w}+2\pi) = e^{ \pm 2\pi i \al } \tpsi^{\pm}(\bar{w}), \hspace{1cm} 
\tpsi^{\pm} ( \bar{w}+ 2\pi \bar{\tau}) = e^{\pm 2\pi i \beta} \tpsi^{\pm}(\bar{w}).
\label{twisted bc fermion}
\end{eqnarray}
The calculations are almost parallel, though include some technical complications,
and the twist parameter $u$ eventually turn out to be identified with the `$u$-parameter' 
of the function 
$\hf^{(*)}_u(\tau,z)$ \eqn{hAppell}.
See \cite{ES-nhP} for the detailed arguments.
Closely related studies including such a twisting based on a different approach 
have been given in \cite{AT,AT3}. 
\end{description}

~

~


\section{Applications to Gepner-like Orbifolds and Moonshine Phenomena}

~


In this section we review the main studies given in \cite{ES-N=4L}.

~

\subsection{Gepner-like Orbifolds describing ALE-spaces}

~

As an interesting application of our previous analyses on $SL(2)/U(1)$, 
let us consider the non-compact extensions of Gepner models \cite{Gepner} describing the Calabi-Yau 
compactifications, initiated by \cite{OV,GKP}. 
Especially, the type II string theory defined on ALE space of the type $G \left(\equiv A_m, \, D_m, \, E_m \right)$ 
is described by the superconformal system expressed  schematically as \cite{OV} 
$$
\mbox{type II}/\mbox{ALE($G$)} \cong 
\left. SU(2)/U(1) \otimes SL(2)/U(1) \right|_{U(1)\msc{-charge} \in \bz},
$$
where $SU(2)/U(1)$, $SL(2)/U(1)$ respectively denotes  the Kazama-Suzuki supercoset theories and 
`$U(1)\mbox{-charge} \in \bz$' means the orbifolding with respect to the total $U(1)_R$-charge 
measured by the total $U(1)_R$-current 
$
J^{\msc{tot}}(z) \equiv J^{SU(2)/U(1)}(z) + J^{SL(2)/U(1)}(z).
$
Note that the superconformal symmetry gets enhanced to $\cN=4$, since the total central charge is equal $\hc =2$ \cite{EOTY}.
It is quite interesting that the type of blown-up singularity of ALE-space, denoted by $G$, is naturally encoded 
into the type of modular invariance of affine $SU(2)$ \cite{CIZ,Kato}.

We shall now focus on the simplest case of $A_{N-1}$, in which the relevant superconformal system is expressed as  
\begin{equation}
\mbox{type II}/\mbox{ALE($A_{N-1}$)} \cong 
\left. \frac{SU(2)_{N-2}}{U(1)} \otimes \frac{SL(2)_{N+2}}{U(1)} \right|_{\bz_N\msc{-orbifold}}, 
\label{A_N-1 model}
\end{equation}
with the total central charge
$$
c = \frac{3(N-2)}{N} + \frac{3(N+2)}{N} =6,
$$
($\hc =2$, in other words).
Here 
the orbifolding for the total $U(1)$-charge reduces to the $\bz_N$-orbifolding
and effectively represented in terms of 
the integral spectral flows, similarly to \cite{EOTY} for the compact Gepner models.

The elliptic genus of 
$SU(2)/U(1)$-sector, or equivalently, the $\cN=2$-minimal model
is given by the well-known formula \cite{Witten-E2};
\begin{align}
\cZ^{\msc{(min)}}(\tau,z) & = \sum_{\ell=0}^{N-2}\, \ch{(\stR)}{\ell,\ell+1}(\tau,z)
\equiv \frac{\th_1\left(\tau, \frac{N-1}{N} z\right)}{\th_1\left(\frac{z}{N}\right)},
\label{cZ min}
\end{align}
where $\ch{(\stR)}{\ell,m}(\tau,z)$ denotes the character of  $\cN=2$ minimal model with level $N-2$ ($\dsp \hc = 1 -\frac{2}{N}$) in 
the $\tR$-sector,  of which expression is given in Appendix A. 
To construct the Gepner-like $\bz_N$-orbifolds 
we also need the `spectrally flowed'  elliptic genus which is again expanded in terms of the minimal characters;
\begin{align}
\cZ^{\msc{(min)}}_{[a,b]}(\tau,z) 
& := (-1)^{a+b+ab}\, s^{\left(\frac{N-2}{2N}\right)}_{a\tau +b} \cdot \cZ^{\msc{(min)}}(\tau,z) 
\nn
& \equiv  \sum_{\ell=0}^{N-2}\, e^{2\pi i \frac{b}{N}\left(\ell+1-a \right)}\, \ch{(\stR)}{\ell,\ell+1-2a}(\tau,z)
\hspace{1cm} \left(\any a,b \in \bz_N \right).
\label{cZ min ab}
\end{align}

For the $SL(2)/U(1)$-sector, on the other hand, we here adopt the vector-like model\footnote
   {When instead taking the axial-like model \eqn{EG cigar formula}, the arguments given below are almost parallel. }, and 
rewrite the elliptic genus \eqn{EG vector-like formula} as $\cZ^{SL(2)/U(1)}(\tau,z)$, that is, 
\begin{align}
\cZ^{SL(2)/U(1)}(\tau,z)
& = 
\frac{\th_1(\tau,z)}{i\eta(\tau)^3} \, 
 \hf^{(N)}\left(\tau, \frac{z}{N}\right)
\nn
& = \sum_{v\in \bz_N}\, \hchid^{(\stR)}(v,0;\tau,z).
\label{cZ vect}
\end{align}
The spectrally flowed one is defined in the manner similar to \eqn{cZ min ab};
\begin{align}
\cZ^{SL(2)/U(1)}_{[a,b]}(\tau,z) 
& := (-1)^{a+b+ab}\, s^{\left(\frac{N+2}{2N}\right)}_{a\tau +b} \cdot \cZ^{SL(2)/U(1)}(\tau,z) 
\nn
& \equiv  \sum_{v=0}^{N-1}\, e^{2\pi i \frac{b}{N}\left(v+a \right)}\, 
\hchid^{(\stR)}(v, a ; \tau,z)
\hspace{1cm} \left(\any a,b \in \bz_N \right).
\label{cZ vect ab}
\end{align}

The elliptic genus of the ALE space of $A_{N-1}$-type is now written as 
\begin{align}
\cZ_{\msc{ALE($A_{N-1}$)}}(\tau,z) 
& =  \frac{1}{N} \sum_{a,b \in \bz_N} \,
\cZ^{\msc{(min)}}_{[a,b]}(\tau,z) 
\, \cZ^{SL(2)/U(1)}_{[a,b]}(\tau,z)
\nn
& \equiv 
\sum_{r=1}^{N-1} \, \sum_{a\in \bz_N} \, \ch{(\stR)}{r-1, r-2a} (\tau,z) \,
\hchid^{(\stR)}(N-r,  a ;\tau,  z)
\nn
& \equiv 
- \frac{\th_1(\tau,z)}{i \eta(\tau)^3}\,
\sum_{r=1}^{N-1} \, \sum_{a\in \bz_N} \, \ch{(\stR)}{r-1, r+2a} (\tau,z) \,
\hF^{(N)}(r,  a ;\tau, - z).
\label{cZ ALE}
\end{align}
In the third line we introduced the function $\hF^{(N)}(v,a; \tau,z)$ defined in \eqn{def hF}, 
and made use of the identity 
\begin{equation}
\hF^{(N)}(v,a;\tau,z) = - \hF^{(N)}(-v, -a;\tau,-z), \hspace{1cm} 
(\any v, a \in \bz_N).
\end{equation}


It is well-known that the elliptic genus of K3-surface is given 
as \cite{EOTY,KYY}
\begin{align}
\cZ_{\msc{K3}}(\tau,z) = 8 \left[\left(\frac{\th_3(\tau,z)}{\th_3(\tau,0)}\right)^2
+ \left(\frac{\th_4(\tau,z)}{\th_4(\tau,0)}\right)^2 + \left(\frac{\th_2(\tau,z)}{\th_2(\tau,0)}\right)^2
\right] \equiv 2 \phi_{0,1}(\tau,z).
\label{cZ K3}
\end{align}
Here, $\phi_{0,1}(\tau,z)$ denotes the standard notation  of (holomorphic) weak Jacobi form of 
weight 0, index 1, which is known to be unique up to normalization. 
Since the ALE-space would be regarded as the `non-compact K3-surface', 
it is natural to express \eqn{cZ ALE} in the form as
\begin{eqnarray}
&& \cZ_{\msc{ALE($A_{N-1}$)}}(\tau, z)  =  
\al \phi_{0,1}(\tau,z)
+ \frac{\th_1(\tau,z)^2}{\eta(\tau)^6} \, \hH^{(N)}(\tau),
\label{def hH N}
\end{eqnarray}
where $\hH^{(N)}(\tau)$ is a {\em non-holomorphic\/} modular form of weight 2.
In fact, one can confirm by straightforward calculations that the R.H.S of \eqn{cZ ALE} 
behaves as a non-holomorphic weak Jacobi form of weight 0 and index 1.
Because of the uniqueness of the holomorphic Jacobi form mentioned above, 
one can readily determine the coefficient $\al$ by evaluating the Witten index of \eqn{cZ ALE};
\begin{equation}
\cZ_{\msc{ALE($A_{N-1}$)}}(\tau, 0) = N-1,
\label{WI case 1}
\end{equation}
leading to $\dsp \al = \frac{N-1}{12}$
(see {\em e.g.} \cite{ES-BH}). 


On the other hand, it is much more difficult to determine the `non-holomorphic part' 
$\hH^{(N)}(\tau)$.
By using the identity 
\begin{align}
&
\sum_{m\in \bz_{2N}}\, \ch{(\stR)}{\ell,m}(\tau,z) \Th{m}{N}\left(\tau , - \frac{2z}{N}\right) = 
- \frac{\th_1(\tau,z)}{i \pi } \oint_{w=0} \frac{dw}{w} \, \frac{\Th{\ell+1}{N}^{[-]}(\tau, 2w) }{\th_1(\tau,2w)} ,
\label{branching ALE}
\end{align}
which is essentially the branching relation of $SU(2)/U(1)$ Kazama-Suzuki supercoset \eqn{branching minimal},
one can derive a simple formula
\begin{equation}
\hH^{(N)} (\tau) = 
\frac{\eta(\tau)^3}{i \pi } \oint_{w=0} \frac{dw}{w} \, 
\frac{\hf^{(N)}(\tau,w) }{i \th_1(\tau,2w)} e^{(N-2) G_2(\tau) w^2}.
\label{hHN}
\end{equation}
where 
$G_2(\tau)$ is the (unnormalized) 2nd Eisenstein series \eqn{def G2}.
See \cite{ES-N=4L} for the detailed computation. 
To summarize, we have obtained 
\begin{align}
\cZ_{\msc{ALE($A_{N-1}$)}}(\tau, z) 
& = \frac{N-1}{12} \phi_{0,1}(\tau,z) + \frac{\th_1(\tau,z)^2}{\eta(\tau)^6} \, \hH^{(N)}(\tau)
\nn
& \equiv \frac{N-1}{12} \phi_{0,1}(\tau,z)
+ \frac{\th_1(\tau,z)^2}{\eta(\tau)^3} 
\frac{1}{i \pi } \oint_{w=0} \frac{dw}{w} \, 
\frac{\hf^{(N)}(\tau,w) }{i \th_1(\tau,2w)} e^{(N-2) G_2(\tau) w^2}.
\label{formula cZ ALE}
\end{align}


Now, what can we say about the non-holomorphic function $\hH^{(N)}(\tau)$?
Physically, it may express 
 the spectrum of $\cN=4$ massive representations that measures the `deviation' from the compact K3-background. 
Mathematically, 
it is indeed a (completed) mock modular form of weight 2. 
Substituting the formula \eqn{hAppell nh-E} into \eqn{formula cZ ALE}, 
we obtain the next expression \cite{ES-N=4L};
\begin{align}
\hH^{(N)}(\tau) & = 
\frac{1}{4\pi^2} \left[ N \widehat{G}_2(\tau) + \frac{\del}{\del w} \sum_{\la \in \La'}\, 
\left.
\frac{e^{-\frac{\pi}{\tau_2} N \left\{\left|\la \right|^2 + 2\bar{\la} w + w^2 \right\}}}{\la+w}
\right|_{w=0}  \right]
\nn
&= \frac{1}{4\pi^2} \left[ N \hG_2(\tau) -  
\sum_{\la\in \La'} \, \frac{e^{- \frac{\pi}{\tau_2} N \left|\la\right|^2 }}{\la^2} 
\left\{
1+ \frac{2 \pi N}{\tau_2} \left|\la\right|^2
\right\} \right],
\label{formula hHN}
\end{align}
where we denoted $\La' \equiv \La - \{ 0\}$ and 
set $\dsp \hG_2(\tau) \equiv G_2(\tau) - \frac{\pi}{\tau_2}$ 
(the `completion of $G_2(\tau)$' \eqn{def hG2}).
Its holomorphic part $H^{(N)}(\tau)$  would be obtained by formally taking the limit 
$\bar{\tau} \, \rightarrow \, -i \infty$ while keeping $\tau$ finite in the expression of \eqn{formula hHN}.
Thus, we guess
\begin{eqnarray}
&& H^{(N)}(\tau)
\sim
\frac{N}{4\pi^2} G_2(\tau) 
+ \frac{1}{4\pi^2} \frac{\del}{\del w} \sum_{\la = m\tau +n \in \La'}\, 
\left.
\frac{q^{N m^2} e^{2\pi i (2N) m w}}{\la+w}
\right|_{w=0} .
\label{guess HN} 
\end{eqnarray}
However, the double series appearing in \eqn{guess HN} does not converge, 
and thus we have to be more careful. 
To this end, 
we introduce the symbol of the `principal value';
\begin{equation}
{\sum_{n\neq 0}}^P \, a_n := \lim_{N\,\rightarrow\, \infty} \sum_{n=1}^N \, \left(a_n + a_{-n}\right), 
\hspace{1cm} 
{\sum_{n\in \bz}}^P \, a_n := a_0 + {\sum_{n\neq 0}}^P \, a_n,
\label{def sumP}
\end{equation}
and the correct expression of $H^{(N)}(\tau)$ should be
\begin{align}
\hspace{-5mm}
H^{(N)}(\tau)
& =
\frac{N}{4\pi^2} G_2(\tau) \left.
+ \frac{1}{4\pi^2} \frac{\del}{\del w} \left[ \sum_{m \neq 0}\, {\sum_{n\in \bz}}^P\, 
\frac{q^{N m^2} e^{2\pi i (2N) m w}}{\la+w}  
+ {\sum_{n \neq 0}}^P \, \frac{1}{w+n}
\right] \right|_{w=0}
\nn
& \equiv 
\frac{N-1}{12} - 2N \sum_{n=1}^{\infty} \, \frac{n q^n}{1-q^n}
+  2 \sum_{m=1}^{\infty} \, q^{Nm^2}  \left[ \frac{q^{m}}{\left(1 -q^m \right)^2}
+  Nm \frac{1+q^m}{1-q^m} \right].
\label{formula HN 1}
\end{align}
To derive the second line, we substituted \eqn{def G2}
as well as familiar identities
\begin{eqnarray*}
\sum_{n=1}^{\infty}\, \frac{1}{n^2} \equiv \zeta(2) = \frac{\pi^2}{6},
\hspace{1cm}
\frac{i}{2\pi} \, {\sum_{n\in \bz}}^P \,\frac{1}{z+n} = \frac{1}{2} + \frac{y}{1-y} = -\frac{1}{2} + \frac{1}{1-y},
\hspace{1cm} \left(y\equiv e^{2\pi i z}\right).
\end{eqnarray*}
Curiously,  this function $H^{(N)}(\tau)$ plays a crucial role in 
the `umbral moonshine' phenomena \cite{umbral1,umbral2,CH,HM,HMN} as was first suggested in \cite{CH}.
More precise statement will be presented in the next subsection. 
It should be also equated to the `second helicity supertrace' (supersymmetric index)
counting the space-time BPS states
computed in \cite{HM,HMN} based on a different string-theoretical construction. 
Indeed, one finds the `number theoretical' formula for this index; 
\begin{align}
& \chi_2^{(k,d)}(\tau) = \left(\frac{k}{d} -d \right) E_2(\tau) - 24 \cF^{(k,d)}_2(\tau),
\label{formula chi2}
\\
&\cF^{(k,d)}_2(\tau) = \left(d \sum_{\stackrel{r,s}{kr> d^2 s >0}} -\frac{k}{d}  \sum_{\stackrel{r,s}{d^2 r> k s >0}}  \right) s q^{rs},
\label{formula Fkd}
\end{align} 
on page 12 of \cite{HMN} (derived in ref.[57] of \cite{HMN}, more precisely),
where $E_2(\tau)$ denotes the normalized 2nd Eisenstein series;
$$
E_2(\tau) \left(\equiv \frac{3}{\pi^2} G_2(\tau) \right) = 1 - 24 \sum_{n=1}^{\infty}\, \frac{n q^n}{1-q^n}.
$$
The precise relation between $H^{(N)}(\tau)$ and the index $\chi_2^{(k,c)}(\tau)$ is written as 
\begin{equation}
H^{(N)}(\tau) = \frac{1}{12} \chi_2^{(N,1)}(\tau).
\label{id HN chi2} 
\end{equation} 
This identity has been proved in \cite{ES-N=4L} (Appendix D or `Addendum') 
by directly evaluating the both sides of \eqn{id HN chi2}.


\newpage


\subsection{`Duality' in $\cN=4$ Liouville Theory and Umbral Moonshine}

~


As is familiar, the type II string theory on an ALE space discussed in the previous subsection 
is also interpreted as the NS5-brane system by T-duality \cite{OV}.
Let us focus on the $\mbox{ALE}(A_{N-1})$-case, or the stack of $N$-NS5 branes equivalently, for simplicity. 
In this picture the world-sheet CFT is identified as the  `CHS-system' \cite{CHS},
which is described by a non-compact boson $\phi$ with the linear dilaton charge $\cQ_{\phi} = \sqrt{\frac{2}{N}}$, 
$SU(2)$-WZW model with level $N-2$, and four Majorana fermions 
at least in a free field  realization. 
This is an $\cN=4$ superconformal system of level 1 ($c=3 \hc =6$) as noticed in the previous subsection.

It is interesting that we have another $\cN=4$ theory described by the same field contents, but with {\em different \/} linear dilaton charge 
$\cQ_{\phi} \equiv - (N-1)\sqrt{\frac{2}{N}} $. 
This second $\cN=4$ system
has the total central charge 
\begin{equation}
c \left(\equiv 3 \hc \right) = \left(1 + 3 \cQ_{\phi}^2\right) + \frac{3(N-2)}{N} + 4 \cdot \frac{1}{2} = 6(N-1).
\nonumber
\end{equation}
In other words, the relevant $\cN=4$ SCA has the level $N-1$ ($N-2$ comes from $SU(2)$-WZW, while the 4 fermions add 1 to the level).


Actually, the same field content {\em without the linear dilaton\/} lead to a simple free field realization of the large $\cN=4$
superconformal algebra, and the modification by including the linear dilaton term turns out to induce 
the small $\cN=4$ SCA in the cases of {\em only\/} two different values of linear dilaton;
one has $\hc =2$, and another has $\hc=2(N-1)$, as mentioned above. 
See \cite{ES-N=4L} for the detailed computation. 
These types of reductions from the large $\cN=4$ to the small $\cN=4$ theories  
have been already found in \cite{STV,Ivanov:1988rt}, and also potentially utilized in \cite{Matsuda} in order
to construct the Feigin-Fuchs representation of $\cN=4$ SCFT.
Thus, one may naturally define the `$\cN=4$ Liouville theory' with $\hc=2(N-1)$
by including a suitable Liouville potential, 
which should be constructed in the manner similar to the screening charges presented in \cite{Matsuda}. 


In the context of string theory on
the NS5-NS1 background, 
the first one $(\hc=2)$
is identified with the world-sheet
CFT for the `short string' sector (or that describing the `Coulomb branch
tube'), while the second one $(\hc=2(N-1))$
corresponds to the `long string' sector (or the `Higgs branch
tube') \cite{SW}. They are expected to be dual to each other from the viewpoints of 
$\mbox{AdS}_3/\mbox{CFT}_2$-duality.


Although the possible Liouville potentials for the $\cN=4$ Liouville theory have very complicated forms,  
we can evaluate its elliptic genus due to the invariance under marginal deformations.  
In \cite{ES-N=4L}, we have shown that the elliptic genus is given as 
\begin{align}
\cZ_{\cN=4 ~ \msc{Liouville}} (\tau,z) & = (-1)^{N-1} \, \widehat{\ch{(\stR)}{0}}(N-1, 0 ;\tau,z)
\nn
& \equiv \frac{2 \th_1(\tau,z)^2}{i\eta(\tau)^3\th_1(\tau,2z)}\, \hf^{(N)}(\tau,z),
\label{cZ N=4L}
\end{align}
where 
$ \widehat{\ch{(\stR)}{0}}(N-1, 0 ;\tau,z)$
denotes the modular completion of $\cN=4$ massless character of level $N-1$, isospin $0$
in the $\tR$-sector\footnote
{
This is actually the unique modular completion of  $\cN=4$ massless characters since they are  independent of the value of  isospin $\ell$,  
as was discussed in \cite{ES-nhP}.
}.


We can also rewrite \eqn{cZ N=4L} as 
\begin{align}
\cZ_{\cN=4 ~ \msc{Liouville}} (\tau,z) 
&= \frac{\th_1(\tau,z)}{i\eta(\tau)^3} \, \sum_{r=1}^{N-1}\,
\sum_{a \in \bz_{N}}\, \ch{(\stR)}{r-1,r+2a}(\tau,z) 
\hF^{(N)}(r, a ; \tau, (N-1)z),
\label{cZ N=4L 2}
\end{align}
by using the identity 
\begin{equation}
2 \frac{\th_1(\tau,z)}{\th_1(\tau,2z)} \, \hf^{(N)} (\tau,z)
= \sum_{\ell=0}^{N-2}\,
\sum_{a \in \bz_{N}}\, \ch{(\stR)}{\ell,\ell+1+2a}(\tau,z) \hF^{(N)}(\ell+1, a ; \tau, (N-1)z) ,
\label{main id hf hF}
\end{equation}
which was proved in \cite{ES-N=4L}.
By comparing \eqn{cZ N=4L 2} with \eqn{cZ ALE}, 
we can  observe a very simple `duality correspondence';
\begin{align}
& \hF^{(N)}(v,a ;\tau,-z) ~ \mbox{\bf for ALE$(A_{N-1})$} 
\nn
& \hspace{2cm}
~ \longleftrightarrow ~ \hF^{(N)}(v,a ;\tau, (N-1)z) ~ \mbox{\bf for $\cN=4$ Liouville of level $N-1$}.
\label{correspondence}
\end{align}
Another useful realization of the duality is given as 
\begin{align}
\cZ_{\msc{ALE}(A_{N-1})}(\tau,z) &= \frac{N-1}{12} \phi_{0,1}(\tau,z) + 
\th_1(\tau,z)^2 \, \frac{1}{2\pi i} \oint_{w=0} \frac{dw}{w}\, 
\frac{e^{(N-2) G_2(\tau) w^2}}{\th_1(\tau,w)^2} \, \cZ_{\cN=4\, \msc{Liouville}}(\tau,w)
\nn
& \equiv  \frac{N-1}{12} \phi_{0,1}(\tau,z) 
+ \frac{\th_1(\tau,z)^2}{\eta(\tau)^6} \, \frac{1}{8\pi^3 i} \oint_{w=0}\frac{dw}{w}\, 
\frac{e^{(N-1)G_2(\tau) w^2}}{\sigma(\tau,w)^2} \, \cZ_{\cN=4\, \msc{Liouville}}(\tau,w),
\label{duality rel}
\end{align}
where we introduced the Weierstrass $\sigma$-function \eqn{w sigma} in the second line.
This is just derived from the identity \eqn{formula cZ ALE}.

~


Now, let us make some comments on the relation with the analyses on 
the `umbral moonshine' \cite{umbral1,umbral2,CH,HM,HMN}. 
In \cite{CH}, the authors studied (the holomorphic part of)
the extension of \eqn{cZ ALE} with general modular coefficients determined by 
the simply-laced root system $X$ corresponding to each Niemeier lattice. 
We have $\mbox{rank} \, X =24$ by definition, and let $N$ be the Coxetor number of $X$.  
A Niemeier lattice is explicitly expressed as  
\begin{equation}
X = \coprod_i \, X_i, \hspace{1cm} \sum_i \, \mbox{rank}\, X_i = 24,
\end{equation}
where each $X_i$ is the irreducible component of root system
possessing the common Coxetor number $N$.

We then define
\begin{align}
\cZ_X^{[\hc=2]} (\tau,z) & :=   \sum_i \, \cZ_{\msc{ALE}(X_i)} (\tau,z)
\nn
& \equiv  
-\frac{\th_1(\tau,z)}{i\eta(\tau)^3} \,
\sum_{r, s=1}^{N-1} \, \sum_{a \in \bz_N} \, \cN^{X}_{r, s} \, 
\ch{(\stR)}{r-1, s+2a} (\tau,z) \,
\hF^{(N)}(s,  a ;\tau, - z)
\label{cZ ALE X} 
\end{align}
where we set 
\begin{equation}
\cN^X_{r,s} \equiv \sum_i\, \cN^{X_i}_{r,s},
\end{equation}
and 
$\cN^{X_i}_{r,s}$ denotes the modular invariant coefficients of $SU(2)_{N-2}$ 
associated to the simply-laced root system $X_i$ \cite{CIZ,Kato}.
One may identify 
$\cZ_{\msc{ALE}(X_i)} (\tau,z)$ as the elliptic genus 
of the ALE space  associated to the simple singularity of the type $X_i$. 
In \cite{CH} it was suggested that the root system 
$X= \coprod_i X_i$ should be identified 
as the geometrical data of various K3-singularities.

Since we assume $\mbox{rank} \, X \left( \equiv \sum_i \, \mbox{rank}\, X_i\right) =24$, 
we can rewrite \eqn{cZ ALE X} as 
\begin{equation}
\cZ_X^{[\hc=2]} (\tau,z) = 2 \phi_{0,1}(\tau,z) 
- \frac{\th_1(\tau,z)^2}{\eta(\tau)^3} \, \hh^X(\tau).
\label{cZ ALE X 2}
\end{equation}
Here the non-holomorphic function 
$\hh^X(\tau)$ is the completion of mock modular form of weight $1/2$ 
which can be evaluated similarly to $\hH^{(N)}(\tau)$ given in 
\eqn{formula cZ ALE}.
Then, the umbral moonshine (the version of \cite{CH}) claims that the `umbral group'\footnote
   {The umbral group is defined as the symmetry group of the Niemeier lattice labeled by $X$ modulo the Weyl group associated to the
 root system $X$ \cite{umbral1,umbral2}.} 
$G_X$ should act on the holomorphic part $h_X(\tau)$ of 
$\hh_X(\tau)$.


For example, let us take $X = A_1^{24}$. 
Then, by comparing the holomorphic parts of both sides, we obtain from \eqn{cZ ALE X 2} 
\begin{align}
\cZ_{\msc{K3}}(\tau,z) = 24 \ch{(\stR)}{0}\left(k=1,0 ; \tau,z \right) 
+ \frac{\th_1(\tau,z)^2}{\eta(\tau)^3} \, h^{X}(\tau),
\end{align} 
where $\ch{(\stR)}{0}\left(k=1,0 ; \tau,z \right) $ denotes the $\cN=4$ massless character of level $k=1$ and isospin $\frac{\ell}{2}=0$.
Notice that  $\cZ_{\msc{K3}}(\tau,z) = 2 \phi_{0,1}(\tau,z)$, $\cZ_X^{[\hc=2]} (\tau,z) = 24 \cZ_{\msc{ALE}(A_1)}(\tau,z)$
hold, and we have the identity 
\begin{align}
\left[ \mbox{hol. part of} ~ \cZ_{\msc{ALE}(A_1)}(\tau,z) \right] = \ch{(\stR)}{0}\left(k=1,0 ; \tau,z \right), 
\end{align}
as shown in \cite{ES-BH}.
Then, the umbral group $G_X$ is no other than the Mathieu group ${\bf M}_{24}$,
and we obtain 
\begin{align}
h^X(\tau) &\equiv  -24 \frac{H^{(2)}(\tau) }{\eta(\tau)^3} 
\nn
&= 
2 q^{-\frac{1}{8}}\, \left[-1+ 45 q + 231 q^2 + 770 q^3 + 2277 q^4 + 5796 q^5 
+ \cdots \right].
\label{Mathieu}
\end{align}
All the numerical coefficients of $q^{n-\frac{1}{8}}$ ($n \geq 1$) in \eqn{Mathieu} are known to be strictly equal the dimensions of
some (reducible, in general) representations of ${\bf M}_{24}$.
This remarkable fact is no other than the `Mathieu moonshine' first discovered by \cite{EOT}.




Let us next consider the type $X$ generalization of \eqn{cZ N=4L 2},
which is related to \eqn{cZ ALE X} via the duality correspondence like \eqn{correspondence}; 
\begin{eqnarray}
\hspace{-1cm}
&&
\cZ_X^{[\hc=2(N-1)]}(\tau,z) : =  
\frac{\th_1(\tau,z)}{i\eta(\tau)^3} \,
\sum_{r, s=1}^{N-1} \, \sum_{a \in \bz_N} \, \cN^X_{r, s} \, 
\ch{(\stR)}{r-1, s+2a} (\tau,z) \,
\hF^{(N)}(s,  a ;\tau, (N-1) z).
\label{cZ N=4L X} 
\end{eqnarray}
Similarly to \eqn{cZ ALE X 2}, the R.H.S of  \eqn{cZ N=4L X}  can be decomposed as  
\begin{eqnarray}
\cZ_X^{[\hc=2(N-1)]}(\tau,z)  & =  & 
 \Phi_{0, N-1}^X(\tau,z) -  \frac{\th_1(\tau,z)^2}{\eta(\tau)^3} \, \sum_{r=1}^{N-1}\, \hh_r^X(\tau) 
\chi^{(N-2)}_{r-1}(\tau,2z)
\nn
& \equiv &  \Phi_{0, N-1}^X(\tau,z) - \frac{2 \th_1(\tau,z)^2}{i \eta(\tau)^3 \th_1(\tau,2z)} \,
\sum_{r=1}^{N-1}\, \hh_r^X(\tau) \Th{r}{N}^{[-]}(\tau,2z).
\label{cZ N=4L X 2}
\end{eqnarray}
In the above expression $ \Phi_{0, N-1}^X(\tau,z)$ is a weak Jacobi form of weight 0, index $N-1$, 
which is holomorphic with respect to $\tau$, 
but generically meromophic with respect to $z$.
$\chi^{(k)}_{\ell}(\tau,z)$ is the affine $SU(2)$ character of level $k$, isospin $\ell/2$,  and 
$\hh^X_r(\tau)$ are the completions of vector valued mock modular forms of weight 1/2.
%
The function $\hh_r^X(\tau) $ are uniquely determined by imposing the  {\em `optimal growth condition'} given in \cite{umbral2};
\begin{equation}
\lim_{\tau\, \rightarrow\, i\infty}\, q^{\frac{1}{4N}} \left| \hh^X_r(\tau) \right| < \infty,  \hspace{1cm} (\any r=1, \ldots, N-1),
\label{optimal growth cond}
\end{equation}
and the umbral group $G_X$ acts on its holomorphic part $h_r^X(\tau) $ \cite{umbral2}.



We here note the duality relation 
which is the natural extension of 
\eqn{duality rel};
\begin{align}
\cZ_X^{[\hc=2]}(\tau,z) 
& =  2 \phi_{0,1}(\tau,z) 
+ \frac{\th_1(\tau,z)^2}{\eta(\tau)^6} \, \frac{1}{8\pi^3 i} \oint_{w=0}\frac{dw}{w}\, 
\frac{e^{(N-1)G_2(\tau) w^2}}{\sigma(\tau,w)^2} \, \cZ_X^{[\hc=2(N-1)]}(\tau,w).
\label{duality rel cZX}
\end{align}
Now, substituting the decompositions \eqn{cZ ALE X 2}, \eqn{cZ N=4L X 2} into the formula
\eqn{duality rel cZX}, we find
\begin{equation}
\hh^X(\tau) = \sum_{r=1}^{N-1} \, \hh^X_r(\tau) \chi^{(N-2)}_{r-1}(\tau,0) 
\equiv \frac{1}{\eta(\tau)^3}  \sum_{r=1}^{N-1} \, \hh^X_r(\tau) S_{r,N}(\tau),
\label{rel hhX}
\end{equation}
where we 
introduced the `unary theta function' \cite{Zwegers}
\begin{equation}
S_{r, N}(\tau) :=  \left. \frac{1}{2\pi i} \del_z \Th{r}{N}(\tau,2z)\right|_{z=0}
\equiv \sum_{n\in r+2N \bz}\, n q^{\frac{n^2}{4N}}.
\label{def S}
\end{equation}
In fact, the contour integral
$
\dsp
\oint \frac{dw}{w} \, \frac{e^{(N-1) G_2(\tau) w^2}}{\sigma(\tau,w)^2}\, \Phi^X_{0,N-1}(\tau,w)
$
has to be a holomorphic modular form of weight 2, and thus vanishes.
This is the duality relation between 
the expansion coefficients of massive representations of 
 $\cZ^{[\hc=2]}_X (\tau,z)$ and  $\cZ^{[\hc=2(N-1)]}_X (\tau,z)$. 
 In the case of Mathieu moonshine ($N=2,X=A^{24}_{1}$) one has the self-dual situation 
 $\widehat{h}^{A^{24}_{1}}(\tau)=\widehat{h}_{r=1}^{A^{24}_{1}}(\tau)$. In general 
holomorphic parts of $\hh_r^X(\tau)$ should reproduce mock modular form of umbral moonshine
on which the umbral group $G_X$ should act \cite{umbral1}.  

In this section, we have discussed how the umbral moonshine can be reproduced from the two $\cN=4$ superconformal systems
that are related by the duality correspondence.  
This could provide a novel duality picture
of moonshine phenomena based on superstring theory, or the $\mbox{AdS}_3/\mbox{CFT}_2$-correspondence; 
one is that of the world-sheet and the other is of the space-time. 

~

~



\noindent
{\bf Acknowledgments}

I thank Profs. Tamiaki Yoneya  and Tadashi Takayanagi so much for offering me a good opportunity 
to publish this review paper about my works in collaboration with Prof. Tohru Eguchi.


\newpage

\section*{Appendix A: ~ Notations and Useful Formulas}

\setcounter{equation}{0}
\def\theequation{A.\arabic{equation}}


~

In Appendix A we summarize the notations adopted in this paper and related useful formulas.
We assume $\tau\equiv \tau_1+i\tau_2$, $\tau_2>0$ and 
 set $q:= e^{2\pi i \tau}$, $y:=e^{2\pi i z}$.


~

\begin{description}

\item[\underline{Theta functions :}]
%
%
 \begin{equation}
 \begin{array}{l}
 \dsp \th_1(\tau,z)=i\sum_{n=-\infty}^{\infty}(-1)^n q^{(n-1/2)^2/2} y^{n-1/2}
  \equiv 2 \sin(\pi z)q^{1/8}\prod_{m=1}^{\infty}
    (1-q^m)(1-yq^m)(1-y^{-1}q^m), \\
 \dsp \th_2(\tau,z)=\sum_{n=-\infty}^{\infty} q^{(n-1/2)^2/2} y^{n-1/2}
  \equiv 2 \cos(\pi z)q^{1/8}\prod_{m=1}^{\infty}
    (1-q^m)(1+yq^m)(1+y^{-1}q^m), \\
 \dsp \th_3(\tau,z)=\sum_{n=-\infty}^{\infty} q^{n^2/2} y^{n}
  \equiv \prod_{m=1}^{\infty}
    (1-q^m)(1+yq^{m-1/2})(1+y^{-1}q^{m-1/2}), \\
 \dsp \th_4(\tau,z)=\sum_{n=-\infty}^{\infty}(-1)^n q^{n^2/2} y^{n}
  \equiv \prod_{m=1}^{\infty}
    (1-q^m)(1-yq^{m-1/2})(1-y^{-1}q^{m-1/2}) .
 \end{array}
\label{th}
 \end{equation}
\begin{eqnarray}
 \Th{m}{k}(\tau,z)&=&\sum_{n=-\infty}^{\infty}
 q^{k(n+\frac{m}{2k})^2}y^{k(n+\frac{m}{2k})} .
 \end{eqnarray}
 We use abbreviations; $\th_i (\tau) \equiv \th_i(\tau, 0)$
 ($\th_1(\tau)\equiv 0$), 
$\Th{m}{k}(\tau) \equiv \Th{m}{k}(\tau,0)$.
 We also set
 \begin{equation}
 \eta(\tau)=q^{1/24}\prod_{n=1}^{\infty}(1-q^n).
 \end{equation}
%
%
The spectral flow properties of theta functions are summarized 
as follows ($\any m,n\in \bz$);
\begin{eqnarray}
 && \th_1(\tau, z+m\tau+n) = (-1)^{m+n} 
q^{-\frac{m^2}{2}} y^{-m} \th_1(\tau,z) , \nn
&& \th_2(\tau, z+m\tau+n) = (-1)^{n} 
q^{-\frac{m^2}{2}} y^{-m} \th_2(\tau,z) , \nn
&& \th_3(\tau, z+m\tau+n) = 
q^{-\frac{m^2}{2}} y^{-m} \th_3(\tau,z) , \nn
&& \th_4(\tau, z+m\tau+n) = (-1)^{m} 
q^{-\frac{m^2}{2}} y^{-m} \th_4(\tau,z) , 
\nn
&& \Th{a}{k}(\tau, 2(z+m\tau+n)) = 
e^{2\pi i n a} 
q^{-k m^2} y^{-2 k m} \Th{a+2km}{k}(\tau,2z).
\label{sflow theta}
\end{eqnarray}

%
%
%
~

The next identity is useful for our calculations;
\begin{align}
&\frac{\th_1(\tau,u+z)}{\th_1(\tau,u)} = \frac{\th_1(\tau,z)}{i\eta(\tau)^3}\,
\sum_{n\in \bz}\, \frac{w^n}{1-yq^n}, 
\nn
& 
\hspace{2cm} \left(y\equiv e^{2\pi i z}, ~ w \equiv e^{2\pi i u}, ~ \tau_2 > 0,  ~ 
0 <  \frac{u_2}{\tau_2} < 1 \right).
\label{theta id}
\end{align}

We introduce the Weierstrass $\sigma$-function;
\begin{eqnarray}
\sigma(\tau,z) & := & 
e^{\frac{1}{2} G_2(\tau) z^2}\,
\frac{\th_1(\tau,z)}{2\pi \eta(\tau)^3}
\nn
& \equiv & z \prod_{\om \in \La'} \, \left(1-\frac{z}{\om} \right)\, 
e^{\frac{z}{\om} + \frac{1}{2} \left(\frac{z}{\om}\right)^2}.
\hspace{1cm} \left(\La' \equiv \La - \{ 0\}\right),
\label{w sigma}
\end{eqnarray}
where $G_2(\tau)$ is the (unnormalized) second Eisenstein series;
\begin{eqnarray}
G_2(\tau) & := & \sum_{n \in \bz-\{0\}}\, \frac{1}{n^2} + \sum_{m\in \bz-\{0\}}\, \sum_{n\in \bz} \, \frac{1}{(m\tau+n)^2}
\nn
& \equiv & \frac{\pi^2}{3} 
\left[1 - 24 \sum_{n=1}^{\infty}\, \frac{n q^n}{1-q^n} 
\right],
\label{def G2}
\end{eqnarray}
It is useful to note the anomalous $S$-transformation formula of $G_2(\tau)$;
\begin{equation}
G_2 \left(- \frac{1}{\tau} \right) = \tau^2 G_2(\tau) -2\pi i \tau.
\label{G2 S}
\end{equation}
We also set
\begin{equation}
\hG_2(\tau) := G_2(\tau) -\frac{\pi}{\tau_2},
\label{def hG2}
\end{equation}
which is a non-holomorphic modular form of weight 2.


~

\item[\underline{Spectral flow operator :}]
(see also \cite{Eich-Zagier})
\begin{eqnarray}
s^{(\kappa)}_{\la} \cdot f(\tau,z) &:= & e^{2\pi i \frac{\kappa}{\tau_2} \la_2 \left(\la + 2z\right) }\,
f(\tau, z+\la)
\nn
& \equiv &q^{\kappa \al^2} y^{2\kappa \al} e^{2\pi i \kappa \al \beta}\,
f(\tau,z+\al \tau+ \beta) ,
\nn
&& 
\hspace{3cm} 
(\la \equiv \al \tau +\beta, ~ \any \al, \beta \in \br).
\label{def sflow op}
\end{eqnarray}

An important property of the spectral flow operator $s^{(\kappa)}_{\la}$ is 
the modular covariance, which precisely means the following:
\\
Assume that $f(\tau,z)$ is an arbitrary function with the modular property;
$$
f(\tau+1,z) = f(\tau,z), \hspace{1cm} 
f\left(-\frac{1}{\tau}, \frac{z}{\tau}\right) = e^{2\pi i \frac{\kappa}{\tau} z^2} \tau^{\al} \, f(\tau,z),
$$
then, we obtain for $\any \la \in \bc$
$$
s^{(\kappa)}_{\la} \cdot f(\tau+1,z) = 
s^{(\kappa)}_{\la} \cdot f(\tau,z), 
\hspace{1cm} 
s^{(\kappa)}_{\frac{\la}{\tau}} \cdot f\left(-\frac{1}{\tau}, \frac{z}{\tau}\right) 
= e^{2\pi i \frac{\kappa}{\tau} z^2} \tau^{\al}\, s^{(\kappa)}_{\la} \cdot f(\tau,z).
$$ 

~

The next `product formula' is also  useful;
\begin{equation}
s^{(\kappa)}_{\la} \cdot s^{(\kappa)}_{\la'} = 
e^{- 2\pi i \frac{\kappa}{\tau_2} \Im (\la \bar{\la'} )} s^{(\kappa)}_{\la+\la'}
= e^{- 4 \pi i \frac{\kappa}{\tau_2} \Im (\la \bar{\la'} )} s^{(\kappa)}_{\la'} \cdot s^{(\kappa)}_{\la},
\label{product sflow op}
\end{equation}
in other words, 
$$
s^{(\kappa)}_{\al\tau+\beta} \cdot s^{(\kappa)}_{\al'\tau+\beta'} = 
e^{-2\pi i \kappa(\al \beta' - \al' \beta)} \, s^{(\kappa)}_{(\al+\al')\tau+(\beta+\beta')}
=  e^{-4\pi i \kappa(\al \beta' - \al' \beta)} \, s^{(\kappa)}_{\al'\tau+\beta'} \cdot s^{(\kappa)}_{\al \tau + \beta}.
$$
We should note that the spectral flow operators do not commute with each other in general.

~

%

~


\item[\underline{Error function} :]
\begin{equation}
\erf(x) := \frac{2}{\sqrt{\pi}} \int_0^{x} e^{-t^2} dt,  \hspace{1cm} (x\in \br)
\label{Erf}
\end{equation}

The next identity is elementary but  useful;
\begin{eqnarray}
&& \sgn(\nu +0) -  \erf(\nu ) 
= \frac{1}{i\pi} 
\int_{\br - i0}\, dp \, \frac{e^{-(p^2 + \nu^2)}}{p-i\nu}.
\hspace{1cm} (\nu \in \br),
\label{id erf}
\end{eqnarray}


~

\item[\underline{weak Jacobi forms} :]

~

The weak Jacobi form \cite{Eich-Zagier} for 
the full modular group $\Gamma(1) \equiv SL(2,\bz)$
with weight $k (\in \bz_{\geq 0})$ and index
$r (\in \frac{1}{2} \bz_{\geq 0})$ 
is defined by the conditions 
\begin{description}
 \item[(i) modularity :] 
\begin{eqnarray}
 && \hspace{-1cm}
\Phi\left(
\frac{a\tau+b}{c\tau+d}, \frac{z}{c\tau+d}
\right) 
= e^{2\pi i r \frac{c z^2}{c\tau+d}} (c\tau+d)^k \, \Phi(\tau,z), ~~~
\any \left(
\begin{array}{cc}
 a& b\\
 c& d
\end{array}
\right) \in \Gamma(1).
\label{modularity J}
\end{eqnarray}
\item[(ii) double quasi-periodicity :] 
\begin{eqnarray}
 \Phi(\tau,z+m\tau+n) = (-1)^{2r(m+n)} q^{-r m^2} y^{-2r m}\, \Phi(\tau,z),
 \hspace{1cm}
 \left(\any m,n \in \bz \right).
\label{sflow J}
\end{eqnarray}
%
\end{description}
In this paper, we shall use this terminology in a broader sense. 
We allow a half integral index $r$, and more crucially,  allow non-holomorphic dependence on $\tau$,
while we keep the holomorphicity with respect to $z$ \footnote
{According to  the original terminology of \cite{Eich-Zagier},
the `weak Jacobi form' of weight $k$ and index $r$ ($k, r \in \bz_{\geq 0}$) means that $\Phi(\tau,z)$ should 
be Fourier expanded as 
$$
\Phi(\tau,z) = \sum_{n\in \bz_{\geq 0}}\, \sum_{\ell \in \bz}\, c(n,\ell) q^n y^{\ell},
$$
in addition to the conditions \eqn{modularity J} and \eqn{sflow J}.
It is called the `Jacobi form' if it further satisfies  the condition: 
$c(n,\ell) =0$ for $\any n, \ell$ s.t.  $4n r -\ell^2 <0$.
}.


~

\item[\underline{Character Formulas for $\cN=2$ Minimal Model} :]
~

The character formulas of 
the level $k$ $\cN=2$ minimal model $(\hat{c}=k/(k+2))$ \cite{Dobrev,RY1}
are described 
as the branching functions of 
the Kazama-Suzuki coset \cite{KS} $\dsp \frac{SU(2)_k\times U(1)_2}{U(1)_{k+2}}$
defined by
\begin{eqnarray}
&& \chi_{\ell}^{(k)}(\tau,w)\Th{s}{2}(\tau,w-z)
=\sum_{\stackrel{m\in \bsz_{2(k+2)}}{\ell+m+s\in 2\bsz}} \chi_m^{\ell,s}
(\tau,z)\Th{m}{k+2}(\tau,w-2z/(k+2))~, \nn
&& \chi^{\ell,s}_m(\tau,z) \equiv  0~, ~~~ \mbox{for $\ell+m+s \in 2\bz+1$}~,
\label{branching minimal}
\end{eqnarray}
where $\chi_{\ell}^{(k)}(\tau,z)$ is the spin $\ell/2$ character of 
$SU(2)_k$;
\begin{eqnarray}
&&\chi^{(k)}_{\ell}(\tau, z) 
=\frac{\Th{\ell+1}{k+2}(\tau,z)-\Th{-\ell-1}{k+2}(\tau,z)}
                        {\Th{1}{2}(\tau,z)-\Th{-1}{2}(\tau,z)}
\equiv \sum_{m \in \bsz_{2k}}\, c^{(k)}_{\ell,m}(\tau)\Th{m}{k}(\tau,z)~.
\label{SU(2) character}
\end{eqnarray}
The branching function $\chi^{\ell,s}_m(\tau,z)$ 
is explicitly calculated as follows;
\begin{equation}
\chi_m^{\ell,s}(\tau,z)=\sum_{r\in \bsz_k}c^{(k)}_{\ell, m-s+4r}(\tau)
\Th{2m+(k+2)(-s+4r)}{2k(k+2)}(\tau,z/(k+2))~.
\end{equation}
Then, the character formulas of unitary representations 
are written as 
\begin{eqnarray}
&& \ch{(\sNS)}{\ell,m}(\tau,z) = \chi^{\ell,0}_m(\tau,z)
+\chi^{\ell,2}_m(\tau,z),  
\nn
&& \ch{(\stNS)}{\ell,m}(\tau,z) = \chi^{\ell,0}_m(\tau,z)
-\chi^{\ell,2}_m(\tau,z),
\nn
&& \ch{(\sR)}{\ell,m}(\tau,z) = \chi^{\ell,1}_m(\tau,z)
+\chi^{\ell,3}_m(\tau,z), 
\nn
&& \ch{(\stR)}{\ell,m}(\tau,z) = \chi^{\ell,1}_m(\tau,z)
-\chi^{\ell,3}_m(\tau,z).
\label{minimal character}
\end{eqnarray}

\end{description}

~

~


\section*{Appendix B:~ Summary of Modular Completions}

\setcounter{equation}{0}
\def\theequation{B.\arabic{equation}}

~


In Appendix B, we summarize the definitions of  modular completions 
of the irreducible and extended characters of $\cN=2$ SCA 
in the $\tR$-sector
given in \cite{ES-NH,orb-ncpart}.
We again assume $\tau \equiv \tau_1+i\tau_2$, $\tau_2>0$
and 
set 
$q \equiv e^{2\pi i \tau}$, $ y \equiv e^{2\pi i z}$, $  w \equiv e^{2\pi i u} \equiv e^{2\pi i (\al \tau + \beta)}$.

~


\noindent
\underline{\bf Modular Completions of Irreducible Characters : }
\begin{align}
\hchd^{(\stR)}(\la,n;\tau,z) &:= \frac{\th_1(\tau,z)}{2\pi \eta(\tau)^3}  
\,\sum_{\nu \in \la + k\bz}\, \left\{ \int_{\br + i(k-0)} dp\, -
\int_{\br-i0} dp \, y q^{n} 
\right\}\,
\frac{ e^{- \pi \tau_2 \frac{p^2+\nu^2}{k}} }{p-i\nu}
\, \frac{\left(y q^n \right)^{\frac{\nu }{k}}}{1-y q^n}\, y^{\frac{2n}{k}} q^{\frac{n^2}{k}} 
\nn
& \equiv  \chd^{(\stR)} (\la ,n;\tau,z)
+  \frac{\th_1(\tau,z)}{2\pi \eta(\tau)^3} \,
\sum_{\nu \in \la  + k\bz}\,
\int_{\br-i0} dp \, \frac{ e^{- \pi \tau_2 \frac{p^2+\nu^2}{k}} }{p-i\nu}
\, \left(y q^n \right)^{\frac{\nu}{k}} y^{\frac{2n}{k}} q^{\frac{n^2}{k}} 
\nn
& \equiv 
(-1)^n \, 
s^{(\frac{\hc}{2})}_{n\tau}\cdot
\hchd^{(\stR)} (\la,0;\tau,z),
\hspace{1cm}
\left(
\la \in \br, ~ n\in \bz \right),
\label{hchd}
\end{align}
and the irreducible character is defined by \cite{Dobrev}
\begin{align}
\chd^{(\stR)}(\la,n;\tau,z) &:= 
\frac{\th_1(\tau,z)}{i\eta(\tau)^3} \, 
\frac{(y q^n)^{\frac{\la}{k}}}{1-y q^n}\,
 y^{\frac{2n}{k}}  q^{\frac{n^2}{k}}\,
\nn
& \equiv (-1)^n \, 
s^{\left(\frac{\hc}{2}\right)}_{n\tau}\cdot \chd^{(\stR)} (\la,0;\tau,z),
\hspace{1cm}
\left(0 \leq \la \leq k , ~ n\in \bz\right).
\label{chd}
\end{align}
Here 
$\chd^{(\stR)} (\la, n;\tau,z) $
($\hchd^{(\stR)} (\la, n;\tau,z) $)
denotes the (modular completion) of the  character associated to the $n$-th spectral flow of 
discrete irrep. generated by the Ramond vacua; 
\begin{equation}
h= \frac{\hc}{8}, ~~~ 
Q = \frac{\la}{k} - \frac{1}{2}, ~~(0\leq \la \leq k) .
\end{equation}

Note that the modular completion $\hchd^{(\stR)}(\la,n)$ has the periodicity 
under $\la\,\rightarrow \, \la + k$, which is obvious 
from the definition \eqn{chd}, while $\chd^{(\stR)}(\la,n)$ does not.

~


\newpage

\noindent
\underline{\bf Modular Completions of Extended Characters : }

~

We assume  $k = N/K$, ($N,K \in \bz_{>0}$), or equivalently, $\dsp \hc = 1+ \frac{2K}{N}$. 
\begin{align}
\hchid^{(\stR)}(v,a;\tau,z) & :=  \sum_{n\in a+ N\bz}\, 
\hchd^{(\stR)}\left(\frac{v}{K},  n;\tau, z \right)
\nn
& \equiv 
 \frac{\th_1(\tau,z)}{2\pi \eta(\tau)^3}
\,\sum_{r \in v + N\bz}\, \sum_{n \in a + N\bz} \left\{ \int_{\br + i(N-0)} dp\, -
\int_{\br-i0} dp \, y q^{n} 
\right\}\,
\frac{ e^{- \pi \tau_2 \frac{p^2+r^2}{NK}} }{p-i r}
\, \frac{\left(y q^n \right)^{\frac{r}{N}}}{1-y q^n}\, y^{\frac{2n}{k}} q^{\frac{n^2}{k}} 
\nn
& \equiv \chid^{(\stR)}( v ,a ;\tau,z)
+  \frac{\th_1(\tau,z)}{2\pi \eta(\tau)^3}
\sum_{r \in v + N\bz}\, \sum_{n \in a + N\bz} 
\, \int_{\br-i0} dp \, \frac{ e^{- \pi \tau_2 \frac{p^2+r^2}{NK}} }{p-ir}
\, \left(y q^n \right)^{\frac{r}{N}} y^{\frac{2n}{k}} q^{\frac{n^2}{k}} ,
\nn
&
\hspace{10cm}
\left(v,a\in \bz_N \right),
\label{hchid}
\end{align}
and 
\begin{align}
\chid^{(\stR)}(v,a;\tau,z) & :=  \sum_{n\in a + N\bz}\, 
\chd^{(\stR)}\left(\frac{v}{K}, n ;\tau,z\right)
\nn
& \equiv 
\frac{\th_1(\tau,z)}{i\eta(\tau)^3} 
\, \sum_{n\in a+N\bz}\, 
\frac{(y q^n)^{\frac{v}{N}}}{1-y q^n}\,
 y^{\frac{2n}{k}}  q^{\frac{n^2}{k}},
\hspace{1cm} 
\left(a \in \bz_N, ~ 0\leq v \leq N-1\right).
\label{chid}
\end{align}
We note that  $\chid^{(\stR)}(v,a;\tau,z) $ is the extended discrete character 
introduced in \cite{ES-L,ES-BH}. 
Again the modular completion $\hchid^{(\stR)}(v,a)$ is periodic 
under $v\, \rightarrow \, v+N$, while 
$\chid^{(\stR)}(v,a)$ is not.

~


The modular and spectral flow properties of $\hchd^{(\stR)}$, $\hchid^{(\stR)}$
are given as follows \cite{ES-NH,orb-ncpart};
\begin{align}
&
\hchd^{(\stR)} \left(\la,n ; \tau+1, z\right)
= e^{2\pi i \frac{n}{k} \left(\la+ n \right)}\,
\hchd^{(\stR)} \left(\la,n ; \tau, z\right),
\label{T hchd}
\\
& 
\hchd^{(\stR)} \left(\la,n ; - \frac{1}{\tau}, \frac{z}{\tau}\right)
= e^{i\pi \frac{\hc}{\tau}z^2}\,
 \frac{1}{k} \int_0^k d\la'  \,\sum_{n' \in \bz}\,
e^{2\pi i \frac{\la \la' - (\la+2n)(\la'+2n')}{2k}}
\, \hchd^{(\stR)} (\la',n';\tau,z),
\nn
&
\label{S hchd}
\\
& \hchd^{(\stR)} (\la,n;\tau,z+r\tau+s) = (-1)^{r+s}  
e^{2\pi i \frac{\la+2n}{k}s}
q^{-\frac{\hc}{2}r^2} y^{-\hc r}\, \hchd^{(\stR)}(\la,n+r;\tau,z),
\nn
& \hspace{12cm} (\any r,s \in \bz),
\label{sflow hchd}
\end{align}
\begin{align}
&
\hchid^{(\stR)} \left(v,a ; \tau+1, z \right)
= e^{2\pi i \frac{a}{N} \left(v+ K a \right)}\,
\hchid^{(\stR)} \left(v,a ; \tau, z \right),
\label{T hchid}
\\
&
\hchid^{(\stR)} \left(v,a ; - \frac{1}{\tau}, \frac{z}{\tau}\right)
= e^{i\pi \frac{\hc}{\tau}z^2}\,\frac{1}{N} \, \sum_{v'=0}^{N-1} \,\sum_{a'\in \bz_N}\,
 e^{2\pi i \frac{vv' - (v+2Ka)(v'+2Ka')}{2NK}}
\, \hchid^{(\stR)} (v',a';\tau,z),
\label{S hchid}
\\
& 
\hchid^{(\stR)} (v,a;\tau,z+r\tau+s) = (-1)^{r+s} e^{2\pi i \frac{v+2Ka}{N}s}
q^{-\frac{\hc}{2}r^2} y^{-\hc r}\, \hchid^{(\stR)}(v,a+r;\tau,z),
\nn
& 
\hspace{12cm} (\any r,s \in \bz).
\label{sflow hchid}
\end{align}

~

We also note the formula for Witten indices;
\begin{align}
 \lim_{z\,\rightarrow\, 0}\, \chd^{(\stR)} (\la,n;\tau,z) & = 
\lim_{z\,\rightarrow\, 0}\, \hchd^{(\stR)} (\la,n;\tau,z) 
= \delta_{n,0} ,
\nn
 \lim_{z\,\rightarrow\, 0}\, \chid^{(\stR)} (v,a;\tau,z) & = 
\lim_{z\,\rightarrow\, 0}\, \hchid^{(\stR)} (v,a;\tau,z ) =  
\delta_{a,0}^{(N)} 
\equiv 
\left\{
\begin{array}{ll}
1 & ~~~ a\equiv 0~ (\mod N) \\
0 & ~~~ \mbox{otherwise}.
\end{array}
\right.
\label{WI}
\end{align}


~


\end{document}